\documentclass{article}

	\usepackage[usenames,dvipsnames,svgnames,x11names]{xcolor}
	
	\newcommand{\ysnoted}[1]{} 
	
	\makeatletter
	\def\mathcolor#1#{\@mathcolor{#1}}
	\def\@mathcolor#1#2#3{%
	  \protect\leavevmode
	  \begingroup
		\color#1{#2}#3%
	  \endgroup
	}
	\makeatother
	
	\usepackage[normalem]{ulem} 

	\usepackage[nocompress]{cite}
	\usepackage{listings}
	
	\lstdefinelanguage{XML}
	{
	basicstyle=\ttfamily\footnotesize,
	  morestring=[b]",
	  moredelim=[s][\bfseries\color{Maroon}]{<}{\ },
	  moredelim=[s][\bfseries\color{Maroon}]{</}{>},
	  moredelim=[l][\bfseries\color{Maroon}]{/>},
	  moredelim=[l][\bfseries\color{Maroon}]{>},
	  morecomment=[s]{<?}{?>},
	  morecomment=[s]{<!--}{-->},
	  commentstyle=\color{gray},
	  stringstyle=\color{blue},
	  identifierstyle=\color{red}
	}
	\usepackage{moreverb}
	
	\usepackage[nounderscore]{syntax}
	
	\usepackage[pdftex]{graphicx}
	\usepackage{subfig}
	
	\graphicspath{{./figures/}}
	\DeclareGraphicsExtensions{.pdf}
	\usepackage{subfig} 
	\usepackage[cmex10]{amsmath}
	\usepackage{amssymb}
	\usepackage{mathtools}
	\usepackage{amsthm}
	\usepackage{amsfonts}
	\usepackage{bbm}

	\newtheorem{lemma}{Lemma}
	\newtheorem{assertion}{Assertion}
	\usepackage{algorithmicx}
	\usepackage[noend]{algpseudocode}
	\usepackage[ruled]{algorithm}
	\definecolor{light-gray}{gray}{0.75}
	\algrenewcommand{\algorithmiccomment}[1]{\hskip3em{{\footnotesize \textcolor{light-gray}{$\blacktriangleright$}}} #1}
	\usepackage{multirow} 
	\usepackage{rotating} 
	\usepackage{booktabs} 
	\usepackage{colortbl} 
	\usepackage{tablefootnote} 
	
	\usepackage[pdftex,colorlinks=true,urlcolor=blue,citecolor=blue]{hyperref}

	\usepackage{xspace}

	\usepackage{enumitem}
	\newcommand{\para}[1]{\noindent\textbf{#1.~}}

	\usepackage[english]{babel}
	\usepackage{blindtext}

	\date{}
	
\begin{document}
	\title{A Partition-centric Distributed Algorithm for \\Identifying Euler Circuits in Large Graphs\footnote{To appear in Proceedings of 5th IEEE International Workshop on High-Performance Big Data, Deep Learning, and Cloud Computing,	In conjunction with The 33rd IEEE International Parallel and Distributed Processing Symposium (IPDPS 2019),
			Rio de Janeiro, Brazil, May 20th, 2019}}
	\author{Siddharth D Jaiswal and Yogesh Simmhan \\
		\emph{Department of Computational and Data Sciences,}\\
		\emph{Indian Institute of Science (IISc), Bangalore 560012, India}\\  
		\emph{Email: siddharthj@IISc.ac.in, simmhan@IISc.ac.in}}

	\maketitle
	
	\begin{abstract}
	Finding the Eulerian circuit in graphs is a classic problem, but inadequately explored for parallel computation. With such cycles finding use in neuroscience and Internet of Things for large graphs, designing a distributed algorithm for finding the Euler circuit is important. Existing parallel algorithms are impractical for commodity clusters and Clouds. We propose a novel partition-centric algorithm to find the Euler circuit, over large graphs partitioned across distributed machines and executed iteratively using a Bulk Synchronous Parallel (BSP) model. The algorithm finds partial paths and cycles within each partition, and refines these into longer paths by recursively merging the partitions. We describe the algorithm, analyze its complexity, validate it on Apache Spark for large graphs, and offer experimental results. We also identify memory bottlenecks in the algorithm and propose an enhanced design to address it.
	\end{abstract}
		
	\section{Introduction}
	\label{sec:intro}
	Finding an \emph{Euler circuit} through a graph is a classic problem in graph theory, and traces its origins to Leonard Euler's paper on finding a walk through the \emph{Seven bridges of K\"{o}nigsberg} while traversing each bridge exactly once~\cite{euler}. The circuit requires that starting at any vertex, we traverse every edge of the graph that is part of the connected component exactly once, and return back to the starting vertex. Finding such circuits is practically useful in transportation and logistics for route planning~\cite{euler-salt,euler-postman}, and for efficient circuit design~\cite{roy2007}.
	
	A sequential algorithm to find the Euler Circuit by Hierholzer~\cite{hierholzer} is the most popular, and has linear time-complexity in the number of edges, $\mathcal{O}(|E|)$. As a result of its simplicity and low time complexity on sequential and shared-memory machines, there has not been the need for any substantial work on parallel algorithms to find the Euler circuit. 
	
	However, graph datasets are growing large in various domains. We are seeing the use of Euler circuits in large graphs from biology for DNA fragment assembly and rendering~\cite{pevzner2001,euler-dna}, and for routing of autonomous vehicles~\cite{euler-robot,euler-uav} that can span large networks over fine-grained spatial regions. There has also been a proliferation of scalable Big Data platforms and abstractions for graph processing, such as GraphLab~\cite{graphlab}, Google Pregel/Apache Giraph\cite{pregel,giraph} and GraphX\cite{graphx}, which operate on commodity clusters and clouds and have democratized access to computing. This motivates the need for a parallel Euler circuit algorithm that can make effective use of commodity resources and such abstractions.
	
	Graph algorithms tend to be memory-bound, and distributed algorithms allow the entire graph to fit in distributed memory~\cite{graphperfanalysis}. There has been some work on parallel Euler circuit algorithms for the PRAM model of computation~\cite{atallah1984,awerbuch1984}. Specifically, Atallah and Vishkin~\cite{atallah1984} propose an algorithm using the CRCW PRAM model that has a time complexity of $\mathcal{O}(log~|V|)$, where $|V|$ is the number of vertices, using $|V + E|$ processors. However, the PRAM model is not practical for modern distributed systems which we target. Makki~\cite{makki} extends a prior centralized algorithm~\cite{evans} to an iterative distributed solution, where at every step, we traverse from a single active vertex along one of its unvisited out-edges. This resembles a contemporary \emph{vertex-centric} programming model like Pregel~\cite{pregel}, but suffers from a high coordination cost of $\mathcal{O}(|E|)$, which is intractable for large graphs on commodity clusters. Other vertex-centric algorithms identify circuits for trees, but not for general graphs~\cite{yan2014}. 

	In this paper, we propose a novel partition-centric distributed algorithm~\cite{think-vertex,goffish,giraph++} to find the Euler circuit for large graphs. We assume that the graph is an Euler graph, i.e., every vertex has an even number of edges~\cite{hierholzer}. The graph is partitioned into connected components placed on different machines. Intuitively, we then concurrently find pairwise edge-disjoint partial paths locally in each partition, and then recursively merge the partial paths in pairs of partitions, to construct the final circuit. Our algorithm attempts to reduce the memory and communication costs. Finding local paths reduces the memory for a single partition on a machine by replacing many edges with a single path. Merging partition-pairs onto a single machine is a form of coarse-grained communication, while allowing paths to be linked together. To the best of our knowledge, this is the first partition-centric distributed algorithm for finding an Euler Circuit. 
	
	We make the following specific contributions in this paper:
	\begin{enumerate}[leftmargin=0.4cm,itemindent=0cm,labelwidth=0.4cm,labelsep=0cm,align=left]
	\item We design a partition-centric algorithm to find the Euler circuit for large graphs on commodity clusters.  We describe the algorithm and its complexity measures, in Sec.~\ref{sec:algo}. 
	\item We implement the algorithm on Apache Spark~\cite{spark}, and present experimental results and analysis for large synthetic graphs, in Sec.~\ref{sec:results}. 
	\item We identify memory bottlenecks in our design and propose improvements, supported by an analytical study, in Sec.~\ref{sec:improved}. 
	\end{enumerate}
	
	Also, we offer background and review related literature in Sec.~\ref{sec:related}, and present our conclusions and future work in Sec.~\ref{sec:conclusion}.
	
	\section{Background and Related Work}
	\label{sec:related}
	
	\subsection{Distributed Graph Processing Platforms}
	
	A number of graph-processing libraries and frameworks exist for shared-memory and \emph{High Performance Computing (HPC) clusters}. These offer efficient data structures and programming models to design graph algorithms. The \emph{Boost Graph Library (BGL)}~\cite{boost} support various graph algorithms, including Euler circuit, on shared-memory systems, with \emph{Parallel BGL}~\cite{parallelboost} offering an MPI-based implementation. Similarly, \emph{CGMgraph}~\cite{cgmgraph} provides an integrated library of parallel graph methods for clusters using MPI, and implements Euler circuits over forests. \emph{Gunrock}~\cite{gunrock} is a data-centric graph processing library for GPUs using the Bulk Synchronous Parallel (BSP) model~\cite{valiant}. However, such platforms designed for HPC and shared-memory machines are not well-suited for commodity clusters and Clouds, which have networks with higher latency, and is the focus of our work.
	
	Distributed graph processing platforms have been developed in the Big Data community, to operate on \emph{commodity clusters and Clouds}~\cite{pregel,graphlab}.
	In particular, \emph{component-centric models of iterative computing} have emerged~\cite{think-vertex}, where graph components of different granularities -- vertex, subgraph, partition -- being the unit of data-parallel computation on the partitioned and distributed graph~\cite{pregel,goffish,giraph++}. 
	These are based on \emph{Google's Pregel} vertex-centric computing model, where users specify the graph algorithm's logic from the perspective of a single vertex, which can update the local state and pass messages to other vertices. This logic executes in parallel on all vertices, and messages generated are delivered in bulk after a global barrier that waits for all vertices to complete execution. This is then repeated iteratively, as a series of \emph{supersteps} using a BSP execution model~\cite{valiant}, and terminates when all vertices ``vote to halt'' and no messages are in flight.
	
	This model of computing eases the development of distributed graph algorithms by avoiding race conditions, exploiting parallelism and retaining the entire graph in distributed memory. Others have extended this to include coarser units of parallelism, such as \emph{partitions}~\cite{giraph++} and \emph{connected-components}~\cite{goffish}. These allow more progress to be made in a single superstep by processing all entities within the partition or subgraph toward local convergence, and reducing the total number of supersteps required for execution. These reduce the barrier-synchronized coordination costs and also the messaging costs, relative to Pregel. We use the \emph{partition-centric computing model} in our algorithm design.
	
	\emph{Apache Giraph}~\cite{giraph} is an open-source implementation of Pregel. The popular \emph{Apache Spark}~\cite{spark} Big Data platform has also been extended as \emph{GraphX}~\cite{graphx} to support a vertex-centric model. We extend Spark to a partition-centric model to implement our algorithm, leveraging the scalable platform benefits of Spark and the algorithmic benefits of this abstraction.
	
	\subsection{Parallel Eulerian Circuit Algorithms}
	There has been active work on algorithms for Euler circuits, with some parallel and distributed approaches as well. \emph{Fleury}~\cite{fleury} proposed one of the first sequential solutions with a time complexity of $\mathcal{O}(|E|^2)$. 
	But \emph{Hierholzer's}~\cite{hierholzer} solution with a complexity of $\mathcal{O}(|E|)$ is the most common. It selects an initial source vertex in the graph and traverses out along an unvisited edge, repeatedly until it returns to this source and there are no unvisited edges out from it. The algorithm runs again from a source vertex having unvisited edges, and this repeats until all edges are visited. Each run identifies an \emph{edge-disjoint cycle}, or a circuit, in the graph, and they each intersect with at least one other circuit at one or more vertices. \emph{Merging} the cycles at these vertices gives a full Euler circuit.

	Parallel algorithms to find the Euler circuit have been examined in detail for the PRAM computing model.  
	\emph{Tarjan and Vishkin}~\cite{tarjan1985} proposed the Euler tour technique as part of identifying the biconnected components of an undirected graph. This was further improved as a parallel algorithm for finding Euler circuits~\cite{atallah1984,awerbuch1984}, which runs in $\mathcal{O}(\log{}|V|)$ time on a CRCW PRAM model, with $|V| + |E|$ processors having access to the whole graph in shared memory. It first transforms the original graph to an auxiliary bipartite graph, constructs a spanning tree through it, identifies an Euler digraph, followed by an Euler circuit, which is then mapped back to the original graph. However, a key limitation of such PRAM algorithms is that their theoretical speedup does not match reality~\cite{anderson1991comparison}, especially on contemporary distributed computing clusters.
	
	\emph{Makki}~\cite{makki} employs a vertex-centric adaptation of Hierholzer's algorithm and traverses the next unvisited edge from the current vertex. It backtracks when visiting a vertex which has only one unvisited edge so as to construct a single walk and not have to later merge edge-disjoint cycles. While this is compatible with the Pregel model and can even be extended to a partition-centric one, it has performance limitations on large graphs. The number of barrier-synchronized supersteps is equal to the number of edges (vertex-centric) or edge cuts between partitions (partition-centric), which is a high \emph{coordination cost}. Further, all but one machine hosting the partition with the active vertex are idle at a time, causing \emph{poor resource utilization}. 
	We address both these limitations.
	
	\emph{Caceres, et al.}~\cite{caceres1997} identify the Euler circuit on a \emph{tree} using the BSP model. The undirected edges of the tree are replaced by a pair of directed edges, and the successor for each edge in the circuit is identified using sorting. This is followed by list-ranking of each vertex in the circuit. On a system with \emph{p} processors, it requires $\mathcal{O}\big(\frac{|V|}{p}\big)$ memory per processor, and takes $\mathcal{O}(\log{p})$ supersteps and $\mathcal{O}\big(\frac{|V|}{p}\big)$ computations per superstep. \emph{Yan, et al.}\cite{yan2014} take a similar vertex-centric approach
	to find the Euler circuit for a tree. It satisfies their Balanced Practical Pregel Algorithm (BPPA) constraints, and takes 2 supersteps to complete. However, neither generalize to a graph.
	
	Lastly, \emph{external-memory algorithms} have been designed to process large graphs that do not fit in main memory, and these have been used to find the Euler circuit on trees as well~\cite{chiang1995external}. However, they have significant I/O bottlenecks.

	In summary, current literature on finding Euler circuits are sequential algorithms, parallel PRAM algorithms limited to theoretical use, have high complexity for large graphs, or work only on trees and not graphs.  
	We address this gap with a distributed algorithm that uses a partition-centric approach, and offers better complexity metrics that is suitable for scaling to large graphs on commodity clusters. 
	
	\section{Partition-centric Euler Tour}
	\label{sec:algo}

	The starting point for our distributed algorithm is an \emph{undirected graph} that has been partitioned into $p$ parts, each present on a different machine. The expectation is that the partitioner has minimized edge cuts between the partitions, and load-balanced the number of vertices in the partitions, using tools like ParHIP~\cite{parhip}. A consequence of this is that each partition is likely to contain one or more large connected components.
	The graph must be \emph{Eulerian}, with all vertices having an even number of (local or remote) edges.
	
	Such an undirected graph, partitioned across $P_1$--$P_4$, is illustrated in Fig.~\ref{fig:graph}.
	\emph{Internal vertices} in each partition (white, e.g., $v_{4},v_{5}$ in $P_2$) are connected exclusively through \emph{local edges} (red solid lines, e.g., $e_{3,4}, e_{4,5}$) to other vertices in the same partition. \emph{Boundary vertices} (yellow, e.g., $v_{3}$ in $P_2$) have at least one \emph{remote edge} (blue dashed lines) to vertices in other partitions (e.g., between $v_{3}$ in $P_2$ and $v_{13}$ in $P_4$). All vertices are given unique labels, and the edges are uniquely labeled by the pair of vertex IDs that they are incident upon (e.g., $e_{3,13}$ between $v_{3}$ and $v_{13}$). As the graph is undirected, $e_{i,j}$ and $e_{j,i}$ are equivalent. One \emph{Euler circuit} in this graph is: 
	\vspace{-0.1in}
	\begin{gather*} 
	(e_{1,2})_{P_1} \mathcolor{blue}{\rightarrow  e_{2,3} \rightarrow} (e_{3,4} \rightarrow e_{4,5} \rightarrow e_{3,5} )_{P_2} \mathcolor{blue}{\rightarrow e_{3,13} \rightarrow} \\
	(e_{12,13} \rightarrow e_{11,12})_{P_4} \mathcolor{blue}{\rightarrow e_{6,11}} \rightarrow (e_{6,7} \rightarrow e_{7,8} \rightarrow e_{8,9})_{P_3} \\ 
	\mathcolor{blue}{\rightarrow e_{9,10} \rightarrow} (e_{10,12} \rightarrow e_{12,14})_{P_4} \mathcolor{blue}{\rightarrow e_{1,14}}
	\end{gather*}
	For convenience, we have used parenthesis to group local edges within a partition, with the partition ID in the subscript. The blue edges not within parenthesis are remote edges. Alternatively, this can be represented as a series of vertex IDs. 
	
	In the rest of this section, we formally define the partitioned-graph and the Euler circuit problem, followed by our proposed algorithm, its formal properties, and a complexity analysis.
	
		\begin{figure}[t!]
			\centering
			\subfloat[Initial partitioned graph]{
				\includegraphics[width=0.65\columnwidth]{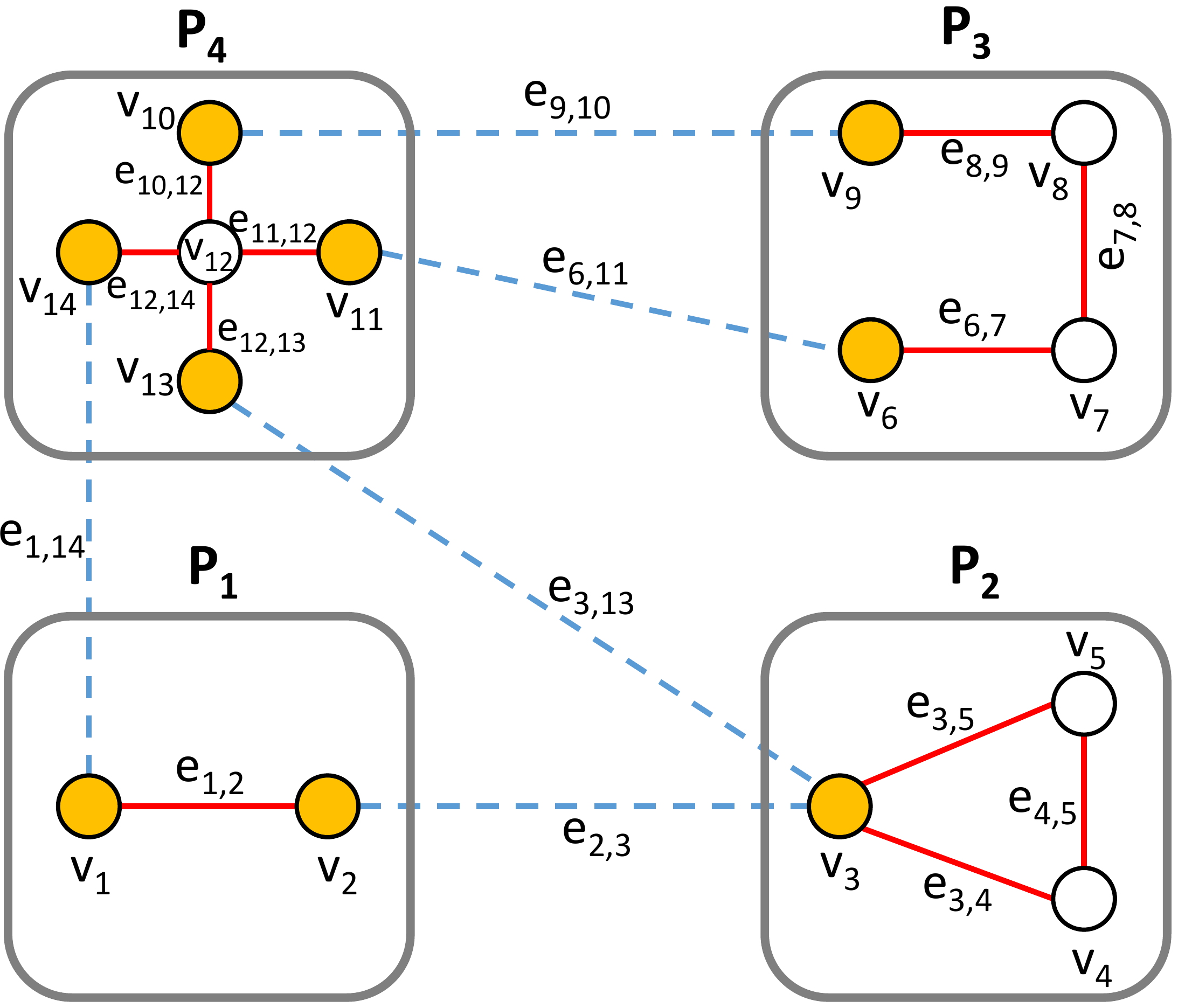}
				\label{fig:graph}}\\
			\vspace{-0.1in}
			\subfloat[Output of Phase 1]{
				\includegraphics[width=0.65\columnwidth]{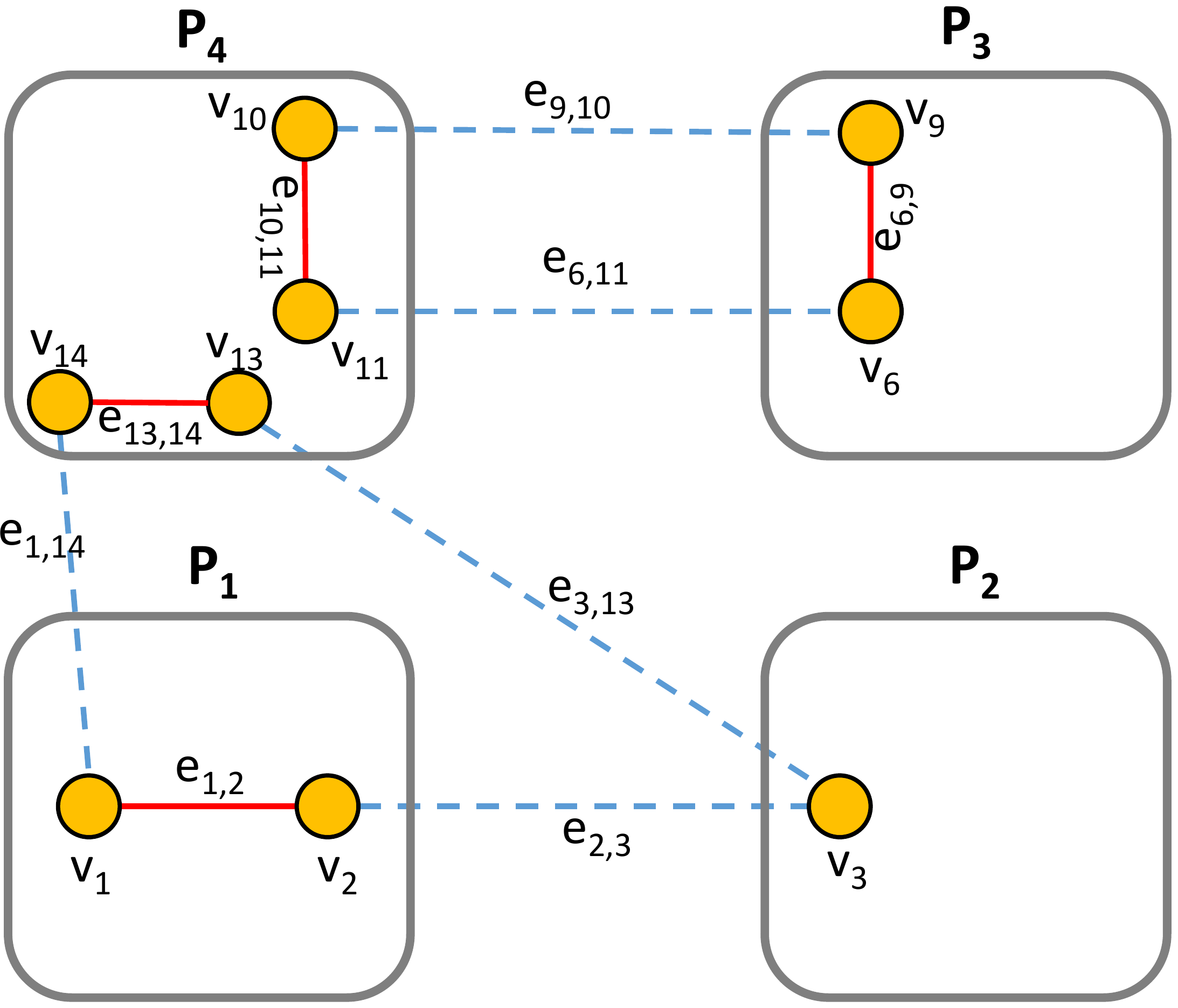}
				\label{fig:phase1}}\\
			\vspace{-0.1in}
			\caption{Incremental phases of finding a partition-centric Euler circuit on a sample graph.}
			\label{fig:completeTour}
		\end{figure}
		
		\begin{figure}[t!]
			\centering
			\ContinuedFloat
			\subfloat[Phase 2 Merging, at Level 1, with $P_2 \leftarrow P_1 \cup P_2$, $P_4 \leftarrow P_3 \cup P_4$] {
				\qquad \includegraphics[width=0.75\columnwidth]{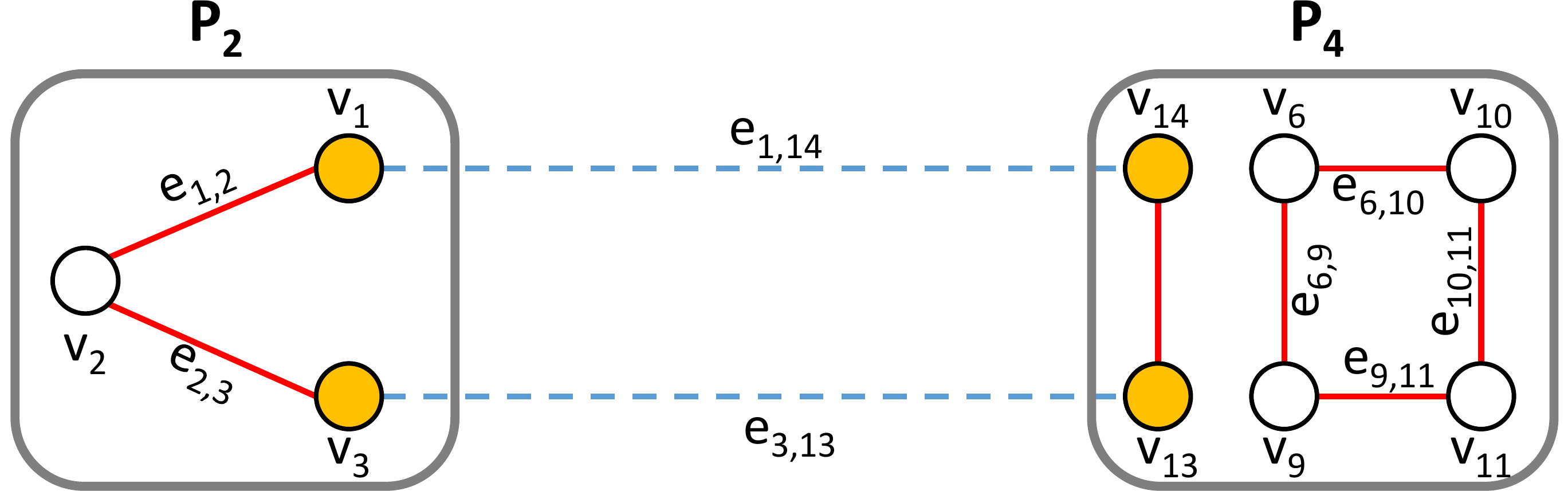} \qquad
				\label{fig:phase2:l1}}\\
			\vspace{-0.1in}
			\subfloat[Output of Phase 1, at Level 1] {
				\qquad \includegraphics[width=0.75\columnwidth]{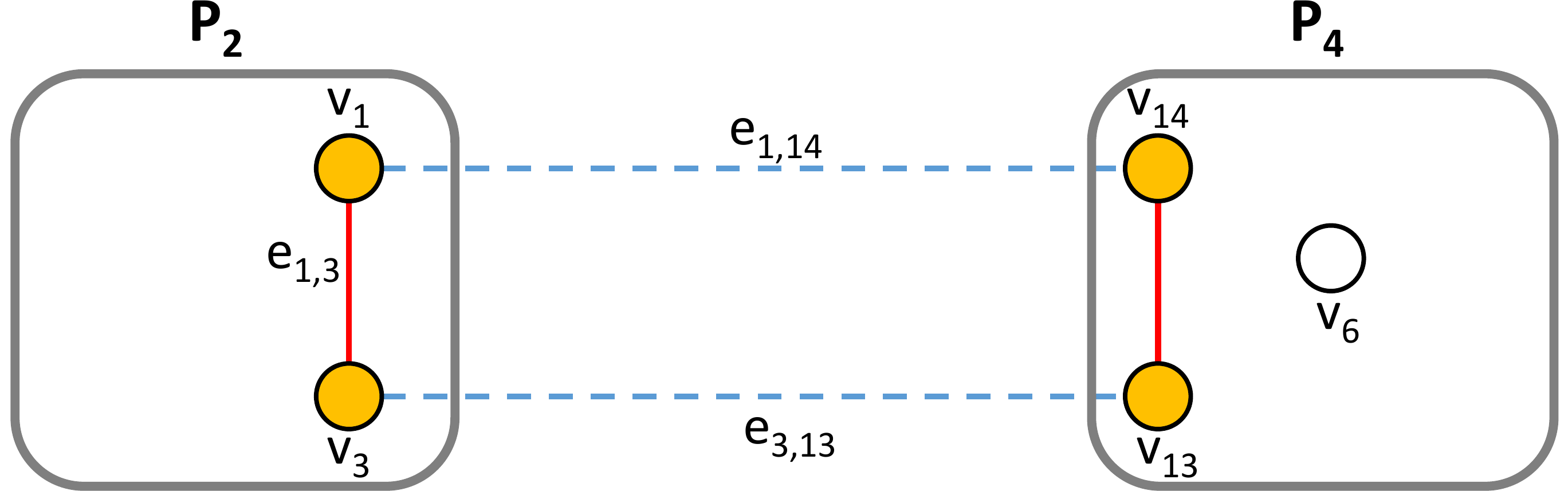} \qquad
				\label{fig:phase1end:l1}}\\
			\vspace{-0.1in}
			\subfloat[Phase 2 Merging, at Level 2, with $P_4 \leftarrow P_2 \cup P_4$]{
				\qquad\qquad\qquad \includegraphics[width=0.25\columnwidth]{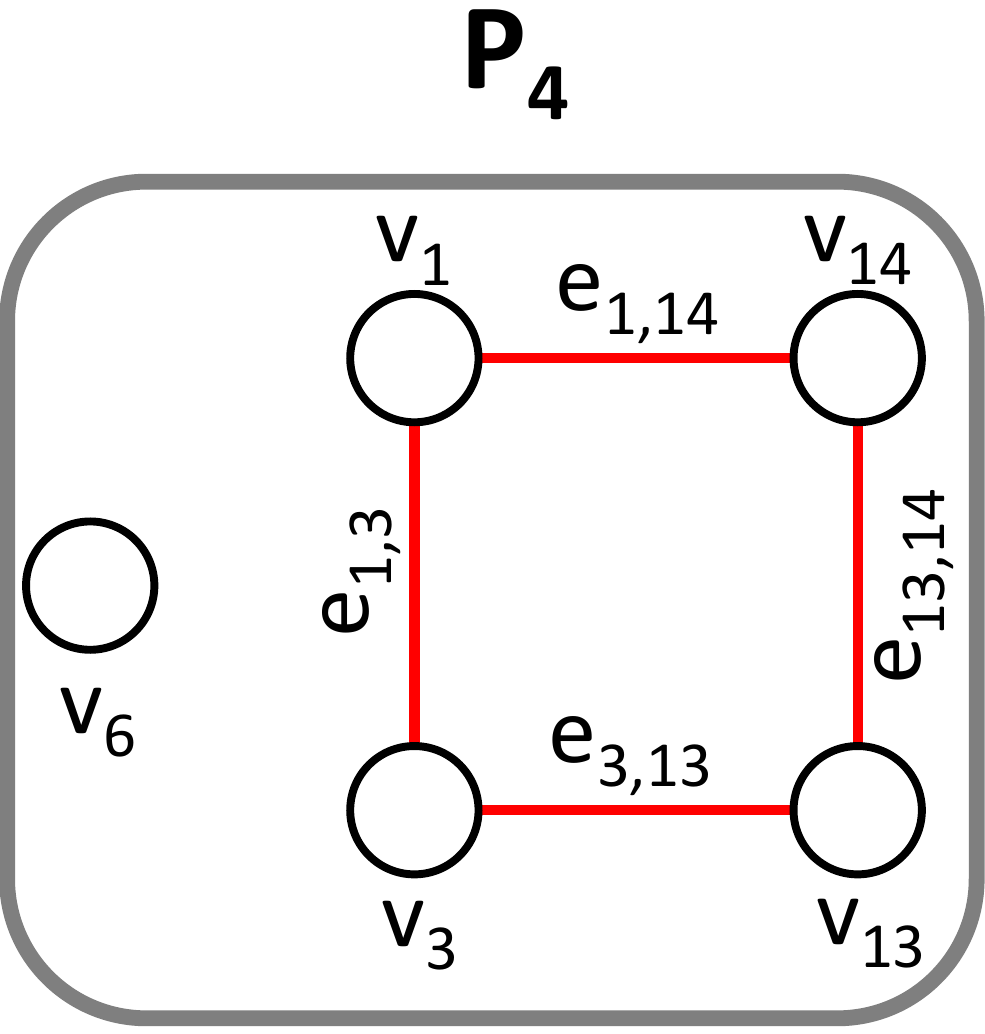} \qquad\qquad\quad
				\label{fig:phase2:l2}}
			\caption{Incremental phases of finding a partition-centric Euler circuit on a sample graph.}
			\label{fig:completeTour}
			\vspace{-0.1in}
		\end{figure}

	\subsection{Preliminaries}\label{prelims}
	An undirected graph is defined as $G = \langle V, E \rangle$, where $v_i \in V$ is the set of vertices and $e_{i,j} \in E \subseteq V \times V$ is the set of edges, where $e_{i,j}$ connects from vertex $v_i$ to $v_j$. We represent an undirected edge with a pair of directed edges, i.e., $e_{i,j} \in E \Leftrightarrow e_{j,i} \in E$.
	
	The \emph{Euler Circuit} on this graph is a \emph{path} $\pi = [ e_{i,j}, e_{j,k}, ... , e_{p, q}, e_{q,i} ] ~\mid~ \forall e_{i,j} \in E ~\exists~ e_{i,j} \in \pi \text{\em , exactly once}$. Here, we require that the path starts \emph{and} ends at the same vertex $v_i \in V$, the sink vertex of an edge is the same as the source vertex of its successor edge, and every edge in $E$ appears exactly once in the path. 
	
	Let $\delta(v_i)$ be the \emph{undirected edge degree} of vertex $v_i$. It has been proved that a graph is \emph{Eulerian}, i.e., has an Euler circuit, if and only if $\forall v_i \in V,~ \delta(v_i) \mod 2 = 0$, i.e., all vertices have an even degree~\cite{euler}.
	
	A graph $G$ partitioned into $n$ parts is given as $\mathbb{G} = \{ P_1,...,P_n \}$, where 
	\vspace{-0.1in}
	\[ P_i = \langle I_i, B_i, L_i, R_i \rangle ~\mid~ I_i, B_i \subseteq V,~~ L_i, R_i \subseteq E \] 
	Here, $I$ and $B$ are the set of \emph{internal} and \emph{boundary vertices} for the partition, while $L$ and $R$ are its set of \emph{local} and \emph{remote edges}. Local edges connect vertices within the same partition while remote edges connect vertices in different partitions, i.e., $l_{p,q} \in L_i \implies v_p, v_q \in (I_i \cup B_i)$ and $r_{p,q} \in R_i \implies v_p \in B_i,~ v_q \in \bigcup_{j \in n \setminus i} B_j$.
	Each vertex in the original graph appears exactly once in one of the partitions, and similarly for each local edge. This is formally stated for vertices as:
	\vspace{-0.1in}
	\[ \bigcup_{i \in n} (I_i \cup B_i) = V \qquad I_i \cap B_i = \varnothing \]
	\vspace{-0.1in}
	\[ \forall i,j \in n, i \neq j,~~ (I_i \cup B_i) \cap (I_j \cup B_j) = \varnothing \]
	and likewise for edges.
	WLOG we treat a partition as a \emph{connected component} to simplify the algorithm design. But it can be generalized to partitions with multiple components, and to the graph itself having multiple components. \ysnoted{``, although the Euler circuit identified on such a disconnected graph will effectively be multiple disjoint circuits on each of these components.''...mixing concepts. we're talking about the partition being disconnected, not the graph}

	We define the \emph{meta-graph} formed from the partitioned graph as
	$\overline{G} =\langle \overline{V}, \overline{E} \rangle$, where the \emph{meta-vertices} $\overline{V} = \{ P_1, ..., P_n\}$ are the set of partitions, and the \emph{meta-edges} $\overline{E} \subseteq \overline{V} \times \overline{V} = \{ m_{i,j} \mid \exists~e_{k,l} \in E, v_{k} \in B_{i},~ v_l \in B_j \}$ denote the existence of at least one edge between their boundary vertices. The \emph{weight of a meta-edge} $m_{i,j}$ is given as $\omega(m_{i,j})$, the count of the edges between all the boundary vertices of those two partitions.

	Further, we classify the boundary vertices for a partition into one of two types: \emph{odd degree boundary vertices ($OB$)} and \emph{even degree boundary vertex ($EB$)}~\footnote{For brevity, we contract these terms and alternatively refer to them as odd vertices (or OB) and even vertices (or EB).}, such that $OB_i \cup EB_i = B_i$ and $OB_i \cap EB_i = \varnothing$. We further have,
	\vspace{-0.1in}
	\[ o_i \in OB \subseteq B \implies \delta_L(o_i) \mod 2 = 1 \] 
	\vspace{-0.2in}
	\[ u_i \in EB \subseteq B \implies \delta_L(u_i) \mod 2 = 0 \]
	where $\delta_L$ and $\delta_R$ give the \emph{local edge degree} and \emph{remote edge degree} for that boundary vertex, respectively. Since we have an Eulerian graph, $\delta_L(v) + \delta_R(v) \mod 2 = 0$. This implies that these odd vertices  
	have an odd numbered remote edge degree as well, and even vertices have an even numbered remote edge degree. In Figure \ref{fig:graph}, the boundary vertices 
	are in yellow, of which $v_{3}$ is an EB with two remote edges, while the rest are OBs. 
	Also, each partition will have an even number of odd vertices, as can be proved using the \emph{Handshaking Lemma}~\cite{euler}. 
	
	\subsection{Approach}\label{sec:approach} 
	\begin{figure}[!t]
		\centering
		\includegraphics[width=0.6\columnwidth]{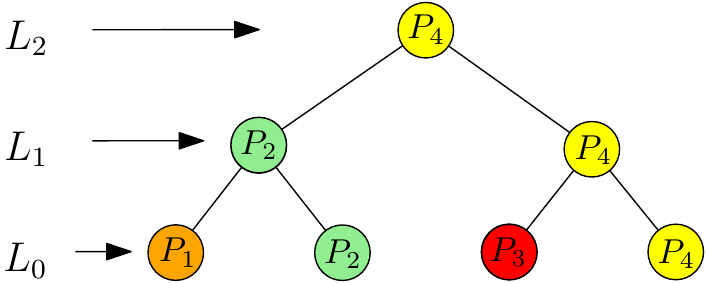}
		\caption{Merge tree for 4 partitions}
		\label{fig:mergeTree}
	\vspace{-0.1in}
	\end{figure}

	We present the high-level idea behind the partition-centric Eulerian circuit, followed by the detailed algorithm, relevant properties and complexity measures. The circuit is calculated in 3 phases. 
	
	\para{Phase 1} Within each partition, we first identify \emph{edge-disjoint maximal local paths} that start and end at an OB through local edges. We then find \emph{edge-disjoint maximal local cycles} using the remaining local edges, which start and end at EBs or internal vertices. This may include trivial tours that start and end at the same even degree boundary vertex, with no edges. 
	This phase is done concurrently on all partitions. 
	
	After this, all local edges $L$ are part of exactly one path or cycle, and all internal vertices $I$ and boundary vertices $B$ are part of one or more paths or cycles. Each path effectively forms a new \emph{coarser} edge, which we call an \emph{OB-pair}, that internally encodes the edges in the path. Likewise, the source/sink boundary or internal vertex encodes the edges in the cycle. The cycles on internal vertices are appended to some OB path or EB cycle.
	So, after Phase 1, the paths and cycles consume and replace all the local edges and internal vertices that are part of them, and \emph{persist the mapping to disk}. This allows the sets $L$ and $I$ to be removed to conserve memory, and only retain the new set of OB-pairs, the boundary vertices $B$, and the remote edges $R$ in memory.
	
	E.g., In Fig.~\ref{fig:graph}, the local path between $v_{6}$ and $v_{9}$ in $P_3$ is $e_{6,7}  \rightarrow e_{7,8} \rightarrow  e_{8,9}$, and this is replaced by the OB-pair edge $e_{6,9}$ in $P_4$, as shown in Fig.~\ref{fig:phase1}. 
	Similarly, the local cycle starting at $v_3$ in $P_2$, $e_{3,4}  \rightarrow e_{4,5} \rightarrow  e_{3,5}$, is consumed and only the boundary vertex $v_3$ is retained in $P_2$.
	
	\para{Phase 2} After Phase 1 completes, we identify pairs of partitions that will be \emph{merged} together. This allows the remote edges between a partition-pair to become local edges of the merged partition, and the corresponding boundary vertices for those edges to become internal vertices (if they have no remote edges to any other partition), or remain boundary vertices. After this, we \emph{rerun Phase 1} on each merged partition to identify additional local paths and cycles that consume these new local edges and internal vertices. As before, this happens concurrently on each merged partition, forming one level. This merge process is repeated recursively, forming a \emph{merge tree} (Fig.~\ref{fig:mergeTree}). We use a greedy strategy to identify partitions to merge that prioritizes partitions with \emph{high meta-edge weight} $\omega$ between them. At the end, only one merged partition remains.

	E.g., Fig.~\ref{fig:phase2:l1} shows the first level of Phase 2 merging performed on the output of Phase 1 in Fig.~\ref{fig:phase1}, with $P_1$ and $P_2$ merged into $P_2$, and $P_3$ and $P_4$ merged into $P_4$. The remote edge $e_{2,3}$ between $P_{1}$ and $P_{2}$ has become a local edge. On running Phase 1 on the merged $P_2$, it would identify $e_{1,2} \rightarrow e_{2,3}$ as a local path between the odd vertices, $v_1$ and $v_3$, which is replaced by the OB-pair, $e_{1,3}$, as seen in Fig.~\ref{fig:phase1end:l1}.
	
	As we can see, Phase 1 reduces the memory usage by the local edges and internal vertices, while Phase 2 increases the memory usage by merging two partitions into one, with the expectation that the memory usage at any point in time is within the limits on a single machine.
	
	\para{Phase 3} At the root partition of the merge tree, we will have a set of internal vertices and cycles (Fig.~\ref{fig:phase2:l2}). We then recursively unroll this coarse-grained information to reconstruct the full Euler circuit, using the book-keeping information persisted to disk. 
	The unrolled path itself can be pushed to disk as it is created.
	Starting at any vertex in this root partition, we unroll the edges of a local cycle that it initiates until we reach a \emph{pivot vertex}, which is a vertex that has multiple paths or cycles intersecting through it.
	We then switch to recursively unrolling edges of a different path or cycle passing through this pivot vertex and created at a lower level, until we return to the pivot vertex, and resume our earlier unrolling.  
	This recursively continues until all edges are emitted. This combines the coarse-grained metadata from the merge tree with the fragments of paths and cycles identified by partitions at different levels into the final circuit, in a single pass.
	
	\subsection{Algorithms} 
	
	\begin{algorithm}[]
	\small
		\caption{Phase 1: Identifying local paths and cycles}\label{algoPhase1}
		\begin{algorithmic}[1]
			\Procedure{doPhase1}{Partition $P_i$}
			\For{$v \in B_{i}$} \Comment{\emph{Init vertices and edges}} \label{algP1:forcheck1s}
			\State $v.visited = false$
			\EndFor \label{algP1:forcheck1f}
			\For{$e \in L_{i}$} \label{algP1:forEdgecheck1s}
			\State $e.visited = false$
			\EndFor \label{algP1:forEdgecheck1f}
			\State {$pathMap[~] = \varnothing$} \label{algP1:initPMap}
			\While {$\exists v.type==\textsc{OB}~\&\&~v.visited==false$} \label{algP1:whilecheck1s} 
			\State $pathMap[~].add(\Call{findEulerPath}{v})$ 
			\EndWhile \label{algP1:whilecheck1f}
			\For {$\forall v.type==\textsc{EB}$} \label{algP1:whilecheck2s} 
			\State $pathMap[~].add(\Call{findEulerPath}{v})$ 
			\EndFor \label{algP1:whilecheck2f}
			\While {$\exists e.visited==false$} \Comment{$e.src \in I_i$} \label{algP1:whilecheck3s}
			\State $ivCycle = \Call{findEulerPath}{e.src}$  
			\State $\Call{mergeInto}{pathMap, ivCycle}$
			\EndWhile \label{algP1:whilecheck3f}
					\Return $pathMap$
			\EndProcedure
		\end{algorithmic}
	\end{algorithm}
	\begin{algorithm}[t!]
	\small
		\caption{Phase 2: Creating merge tree from meta-graph} \label{algoMergeSeq}
		\begin{algorithmic}[1]
			\Procedure{generateMergeTree}{$\overline{G}$}
			\State $mergeTree = \varnothing$ \label{algMerge:initstart}
			\State $ l = 0$ \Comment{\emph{Current level}}
			\State $\overline{G}_l \leftarrow \overline{G}$ \Comment{\emph{Initialize meta-graph at level 0}} \label{algMerge:initend}
			\While {$|\overline{V}_l| > 1$} \label{algMerge:loopstart}
			\State $edges = \Call{maximalMatching}{\overline{E}_l}$
			\State $mergeTree.add(l,edges)$ \hspace{-0.25in}\Comment{\emph{Add partitions incident on each edge as siblings in the tree at level $l$}}
			\State $l \leftarrow l + 1$ \Comment{\emph{Increment level}}
			\State $\overline{G}_{l} = \Call{rebuildMetaGraph}{\overline{G}_{l-1}, edges}$
			\EndWhile \label{algMerge:loopend}
					\Return $mergeTree$
			\EndProcedure
		\end{algorithmic}
	\end{algorithm}
	
	We describe the algorithms for the first two key phases here. We omit a detailed algorithm for Phase 3, for brevity.
	
	\subsubsection{Phase 1} Alg.~\ref{algoPhase1} will be executed concurrently in a partition-centric manner on all the partitions of the initial graph, and on each merged partition after Phase 2. We initialize all boundary vertices and local edges to be unvisited in lines~\ref{algP1:forcheck1s}--\ref{algP1:forEdgecheck1f}, and create an empty $pathMap$ structure in line~\ref{algP1:initPMap}, which will have the local paths and cycles found in this partition. 
	We then call \textsc{findEulerPath} for all unvisited odd degree boundary vertices, in lines~\ref{algP1:whilecheck1s}--\ref{algP1:whilecheck1f}. This procedure navigates along unvisited edges, marking them and their incident vertices as visited, and returns a maximal path. When starting with an OB vertex, this path will end at another OB vertex with no remaining unvisited edges, as we show later in Lemma~\ref{lemma:odbv}. If we have $2n$ OB vertices, this will find exactly $n$ paths.
	
	Once all OB paths are exhausted, we find cycles starting at EBs using \textsc{findEulerPath}. Here, we traverse from every EB exactly once, as shown in lines \ref{algP1:whilecheck2s}--\ref{algP1:whilecheck2f}, and include even a trivial singleton path with zero edges. Later, Lemma~\ref{lemma:edbv} shows that such a maximal path will result in a cycle. So, if we have $m$ EB vertices, this will find exactly $m$ cycles or singletons. 
	
	Lastly, if there still remain any unvisited edges, then either of their incident vertices must be an internal vertex. We initiate a maximal traversal from this vertex, in lines \ref{algP1:whilecheck3s}--\ref{algP1:whilecheck3f}, which will form a cycle, $ivCycle$ (Lemma~\ref{lemma:edbv}). 
	Further, this cycle will intersect with one of the existing OB paths or EB cycles in this partition (Lemma~\ref{lemma:internaladdedtoothers}). \textsc{mergeInto} combines the internal vertex cycle into one of these prior paths/cycles, at a vertex at which they intersect, thus extending the prior ones.
	
	The $pathMap$ assigns a unique \emph{Path ID} to each cycle or path added to it, its \emph{type} (OB path or EB cycle), the \emph{source} and optionally (if it is a path) the \emph{sink boundary vertex ID}, and the \emph{remote edges} for the boundary vertices. The actual vertices and edges in the path/cycle can be persisted to disk. The $pathMap$ is adequate to proceed to the next phase, and other data structures of the partition can be removed. 
		
	\subsubsection{Phase 2} This phase first identifies the pairs of partitions that will be merged recursively, starting with the initial partitions of the graph, until just a single merged partition remains. Alg.~\ref{algoMergeSeq} builds this \emph{merge tree} statically on 1 machine, using just the \emph{meta-graph} -- the list of partition IDs as meta-vertices, their meta-edge connectivity, and their edge weights -- 
	which is small in size, $\mathcal{O}(n(n-1))$, with $n$ partitions and a maximum of $n(n-1)$ meta-edges between them.
	
	Alg.~\ref{algoMergeSeq} uses a greedy approach, with the intuition that partitions with the most edges between them should be merged first as it allows for the consumption of more local edges in the next level's execution of Phase 1. In lines \ref{algMerge:initstart}--\ref{algMerge:initend}, we initialize an empty $mergeTree$, set the level as $0$, and set the meta-graph at this level as the original meta-graph with all partitions.  
	In lines~\ref{algMerge:loopstart}--\ref{algMerge:loopend}, we build one level of the merge tree for each iteration, until only one partition remains. For each level, we find a \emph{maximal matching} of meta-edges over the meta-graph such that all meta-vertices appear exactly once on the selected meta-edges (unless there are an odd number of meta-vertices, in which case one can be skipped), and the sum of the weights of the meta-edges is the highest. This uses a greedy algorithm, \textsc{maximalMatching}, which sorts the meta-edges in descending order of their weights and greedily selects them while ensuring none share a common meta-vertex.
	
	The pairs of meta-vertices (partitions) on each selected edge form siblings in the tree at level $l$, with the parent being one of the two partitions (e.g., the one with a larger partition ID) that is merged into at the next level. We then rebuild the meta-graph for the next level using the merged partitions as the meta-vertices, and repeat this process.
	
	The merge tree for the graph in Fig.~\ref{fig:graph} is shown in Fig.~\ref{fig:mergeTree}. Here, the edge weights between $P_3$ and $P_4$ is the highest, and they are selected for merging at level $0$. That leaves $P_1$ and $P_2$ as the remaining partitions to be merged. We pick $P_2$ and $P_4$ as the parents. In the next level $1$, these two merged partitions are merged again, and into a single partition $P_4$. 
	
	While the merge tree is constructed at the start, the actual merging of two partitions is done after executing Phase 1 on them at each level. This involves transferring the $pathMap$ from the child partitions to the parent, converting the remote edges between the children to local edges, and introducing relevant local and boundary vertices. Once merged, we invoke Phase 1 on each merged partition, in parallel, for this level.

	\subsection{Lemmas} 
	\label{lemmas}
	\begin{assertion}\label{assert:ph1order}
		It is necessary to identify a maximal local paths starting and ending at odd degree boundary vertices, and only then the cycles starting and returning at even degree boundary vertices and at internal vertices, within a partition.
	\end{assertion} 
	\vspace{-0.1in}
	\begin{proof}[\unskip\nopunct]Initiating a traversal from a even internal or boundary vertex is not guaranteed to form a local cycle and return back to the same vertex because an edge it takes may be the last local edge to an odd degree boundary vertex. Hence, by starting traversals at an odd vertex to find paths, and repeating this till all the odd edges are consumed, we ensure that only vertices with an even unvisited degree remain. 
	\end{proof}
	\vspace{-0.1in}
	E.g., in Fig.~\ref{fig:graph}, if we first initiate a traversal at the internal vertex $v_{8}$ of $P_3$, it will terminate at one of the two odd vertices, $v_{9}$ or $v_{6}$, forming a path and not a cycle. \ysnoted{This assertion has to hold for the subsequent lemmas to be true.}
	
	\begin{lemma}\label{lemma:odbv}
		A maximal local path which starts at an odd degree boundary vertex will always end at another odd degree boundary vertex. 
	\end{lemma}
	\vspace{-0.1in}
	\begin{proof}[\unskip\nopunct]
	From \S~\ref{prelims}, we have $\forall~v_p \in I_{i} \cup E_i,~~ \implies \delta_{L}(v_p)~mod~2 = 0 $. Therefore every edge entering an even degree internal or boundary vertex will have a corresponding edge available for leaving that vertex. Thus a path starting at some odd degree boundary vertex and traversing through even vertices will be able to leave those vertices, and can end only at another odd degree boundary vertex which does not have any other unvisited edges. If there are $2n$ odd vertices, there will be exactly $n$ edge-disjoint paths between pairs of them.
	\end{proof} 
	\vspace{-0.1in}
	\begin{lemma}\label{lemma:edbv}
	A local traversal that starts at an even degree boundary vertex or an internal vertex will always end at the same vertex, thus completing a cycle.
	\end{lemma}
	\vspace{-0.1in}
	\begin{proof}[\unskip\nopunct]
	Since all the paths from odd vertices have been taken and their odd edges visited, the remaining vertices will have an even number of unvisited edges, i.e., the internal or even degree boundary vertices and the unvisited edges form an Eulerian graph.   
	Thus any maximal path starting at an even or internal vertex will return back to the same vertex. 
	\end{proof}
	\vspace{-0.1in}
	\begin{lemma}\label{lemma:internaladdedtoothers}
		At least one vertex in any cycle starting at an internal vertex must also be part of a previous odd vertex path or even vertex cycle in the same partition.
	\end{lemma}
	\vspace{-0.1in}
	\begin{proof}[\unskip\nopunct]
	Since partitions form a connected-component,  
	there exists a local path between all pairs of vertices. 
	All maximal local paths starting and ending at odd vertices, and maximal cycles starting and ending at the same even boundary vertex are already taken. So, it follows that a maximal path starting at an internal vertex must pass through a vertex that is part of a prior path or cycle. 
	\end{proof}
	\vspace{-0.1in}
	A \textit{pivot vertex} 
	is one that is present on \emph{more than one} odd degree path, even degree cycle or an internal cycle. In Phase 1, we use these pivot vertices between internal cycles and the odd path or even cycle to merge the former into the latter. In Phase 3, these points of intersection between the various fragments help recursively reconstitute the complete Euler circuit.
	
	\subsection{Complexity Measures}
	\label{complexity}
	There are three costs associated with the distributed partition-centric approach that we have proposed. The \emph{coordination cost} is the total number of iterations (supersteps) that the BSP execution takes. Phase 1 executes on all partitions at a level in \emph{one superstep, in parallel,} and Phase 2's Alg.~\ref{algoMergeSeq} constructs a \emph{full} binary merge tree. So the coordination complexity is the height of the tree, given by $\lceil log(n) \rceil + 1$, where $n$ is the number of partitions in the original graph.
	
	The \emph{computation complexity} is dominated by the Phase 1 algorithm. The complexity of Alg.~\ref{algoPhase1}  executing on each partition is the time to initiate traversals from each boundary vertex and, in the worst case, every internal vertex, and the time to visit every local edge once. This is $\mathcal{O}(|B_i| + |I_i| + |L_i|)$, for a partition $P_i$. This applies to leaf and merged partitions. For each level, all partitions execute concurrently, and so this is bound by the largest partition in a level, and for levels. 
	
	Lastly, the \emph{communication cost} is contributed by the merging of pairs of partitions as part of Phase 2, at each level. The $pathMap$ output from Phase 1 for a partition is sent to the other partition it merges into. The odd and even boundary vertices, and their remote edges form a bulk of this. So the communication complexity is $\mathcal{O}(|B_i| + |R_i|)$, for partition $P_i$ merging into another. Since data transfers happen in parallel at a level, this is bound by the largest partition in a level.
	
	\section{Experiments}
	\label{sec:results}
	\subsection{Implementation}
	\begin{figure}[t]
	\vspace{-0.1in}
		\centering
		\subfloat[Phase 1 on Level 0\label{fig:1b}] {\includegraphics[width=0.33\columnwidth]{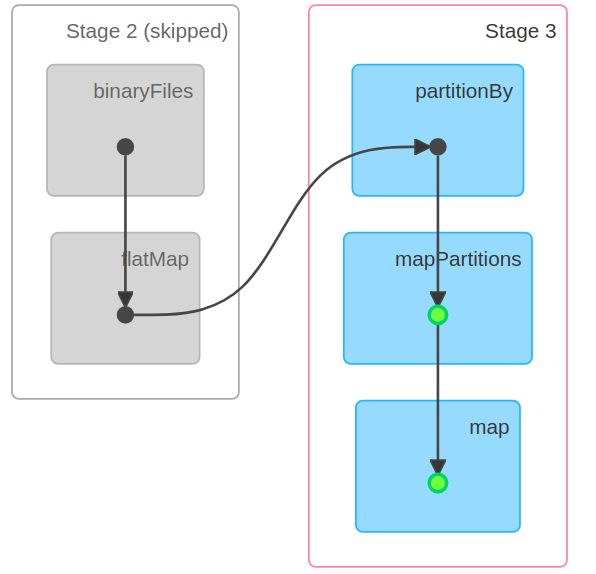}}~~
		\subfloat[Phase 2 Merge, and Phase 1 on Level 1\label{fig:1c}]{\includegraphics[width=0.51\columnwidth]{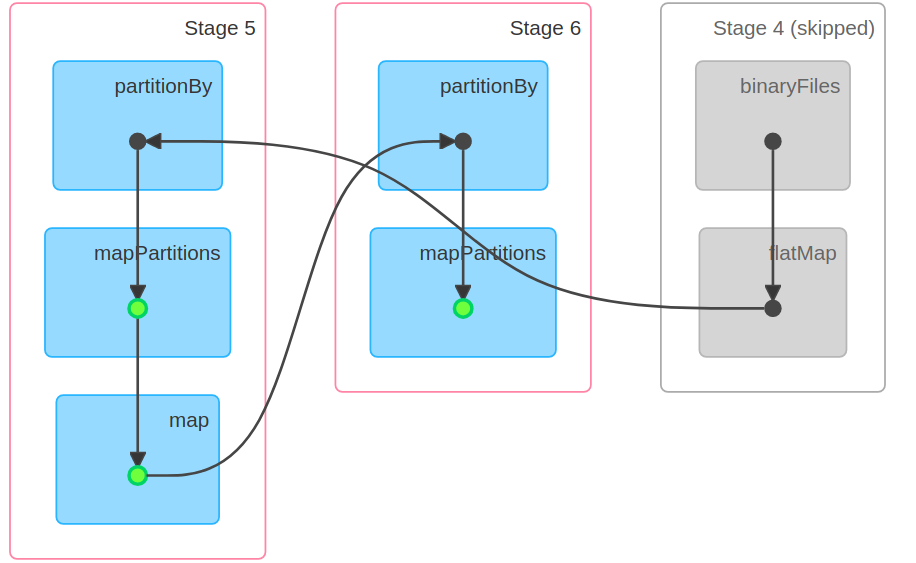}}
		\caption{Stages of execution of algorithms in Apache Spark} \label{fig:sparkdag}
	\vspace{-0.1in}
	\end{figure}
	
	We implement the Phase 1 and 2 algorithms on \emph{Apache Spark 2.2}~\cite{spark} with Java 8, using a partition-centric approach~\footnote{https://github.com/dream-lab/euler-circuit}. The implementation of Phase 3, which can be done sequentially, is left to future work. Since Spark does not naturally expose a partition-centric model, we model all partitions in one level as a single \emph{Resilient Distributed Dataset (RDD)}, with each of our partitions mapping to a Spark partition in the RDD. Phase 1 at a level is designed as a set of \emph{transformations} on the RDD to form a $pathMap$ per partition, and Phase 2 generates a new RDD for the next level by merging partition pairs, based on the merge tree computed offline. This repeats iteratively till an RDD with just one partition remains. Fig.~\ref{fig:sparkdag} illustrates the Spark DAG that executes for levels 0 and 1 of a graph, as captured from the Spark UI.
	
	\subsection{Experiment Setup}
	
	We run the application on a Spark cluster deployed on $8$ Microsoft Azure \texttt{E8s v3} Virtual Machines (VMs), with each having 8~cores of Intel Xeon E5-2673 at 2.3~GHz, 64~GB of RAM, and 128~GB of local storage.  
	Each Spark \emph{executor} runs on a separate VM,
	has access to 45~GB of RAM, and \emph{one executor is assigned to each partition} in the input graph.
	
	\begin{figure}[!t]
		\centering
		\includegraphics[width=0.7\columnwidth,trim={2.5cm 0 2.5cm 3.5cm},clip]{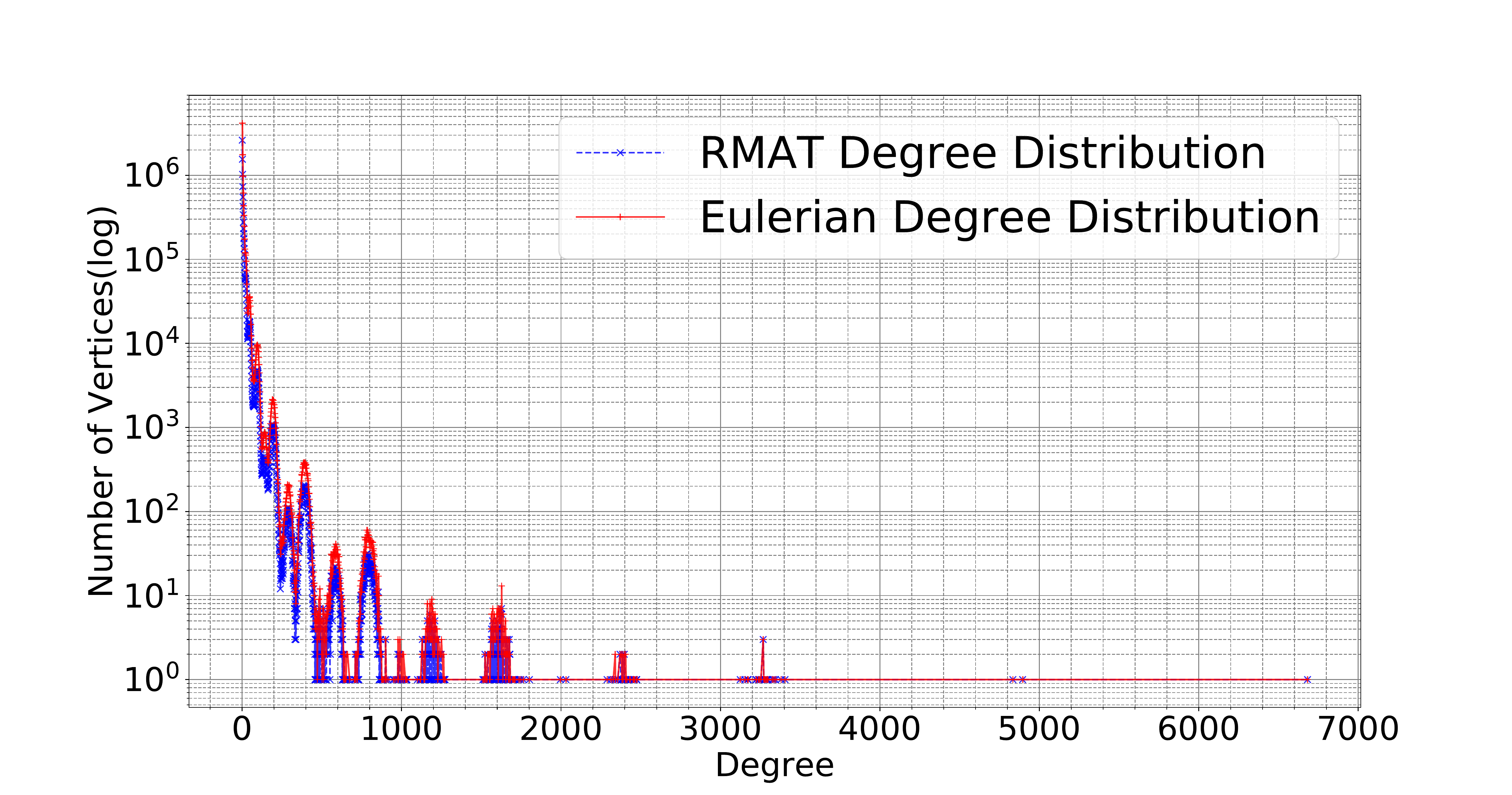}
		\caption{\emph{Edge degree distribution} for the original (Blue) and modified Eulerian graphs (Red), for $10M$ vertices/$50M$ edges.}
		\label{plt:degdist10m}
	\vspace{-0.05in}
	\end{figure}
	
	\begin{table}[t!]
	\setlength{\tabcolsep}{3pt}
		\begin{center}
			\caption{Characteristics of Input Eulerian graphs used}
			\label{tab:dataset}
			\begin{tabular}{c|c c|c c c c}
				\hline
				\textbf{Graph} & $|V|$ & $|E|$ & $\sum_{i \in n}|B_i|$ & {\bf Parts} $(n)$& $\sum{\frac{|R_{i}|}{|E|}}\%$ & $|V_i|$~{\bf Imbal.\%} 
				\\
				\hline\hline
				G20/P2 & 20M & 212M & 13M & 2 & 38\% &19\%\\
				\hline 
				G30/P3 & 30M & 318M & 22M & 3 & 49\% &48\%\\
				\hline
				G40/P4 & 40M & 423M & 23M & 4 & 59\% &46\%\\
				\hline
				G40/P8 & 40M & 423M & 33M & 8 & 70\% &41\%\\
				\hline
				G50/P8 & 49M & 529M & 40M & 8 & 70\% &63\%\\
				\hline
			\end{tabular}
		\end{center}
	\vspace{-0.25in}
	\end{table}
	
	The input we expect is an Eulerian graph, i.e., each vertex has an even edge degree. But there are no well-known tools for generating such graphs. So we first generate a \emph{powerlaw graph} using the \emph{parallel RMAT tool}~\cite{parmat}, and develop a custom tool to add additional edges between vertices that have an odd degree, to make the graph Eulerian. The tool ensures that the edge degree distribution of the modified graph closely matches the original graph, as illustrated in Fig.~\ref{plt:degdist10m}. In practice, the extra edges added is $\approx 5\%$.
	Subsequently, we use the \emph{ParHip tool}~\cite{parhip} to partition these graphs. Table~\ref{tab:dataset} lists the graphs that we generate, and counts of their vertices, edges, boundary vertices, partition, edge-cut fraction, and peak vertex imbalance across partitions, given as \ysnoted{verify} $\max_i{\left| \frac{|V| - n \times |V_{i}|}{|V|} \right|}$.  
	The bi-directed edge counts listed are twice the number of undirected edges. We run RMAT with its default settings, and an average undirected edge degree of $5$.
 
	\subsection{Results}
	
\begin{figure}[!t]
		\centering
		\includegraphics[width=0.75\columnwidth]{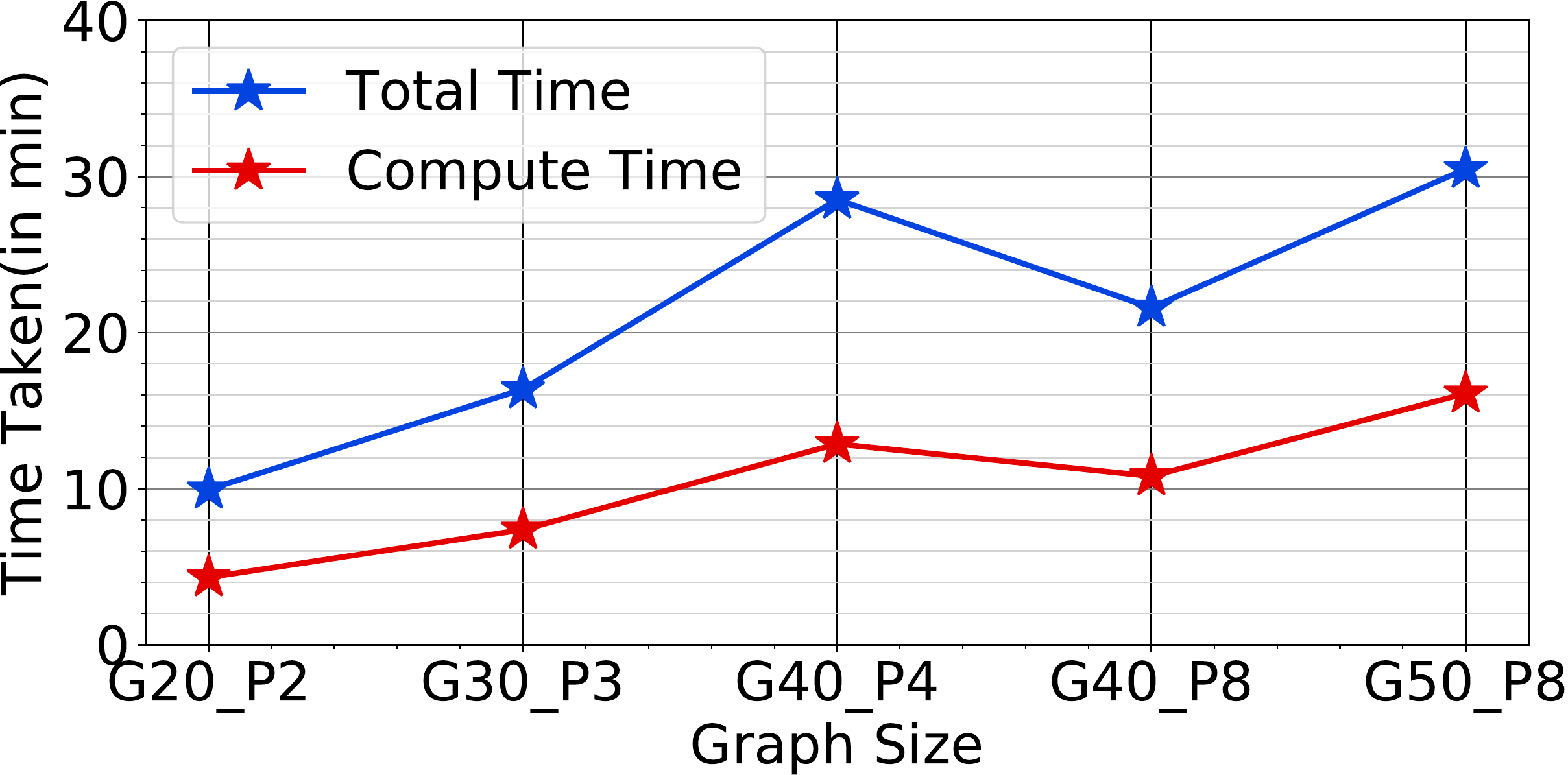}
		\caption{\emph{Total} and \emph{user compute times} for each graph}
		\label{plt:runtime}
	\vspace{-0.1in}
	\end{figure}

	We examine the \textbf{total time} taken to execute our Eulerian circuit Spark application for the 5 candidate graphs, as reported in the Spark logs. This time in minutes (blue line) is shown in the Y axis of Fig.~\ref{plt:runtime} for the different graph sizes and with their partition counts in the X axis. We also plot the \textbf{user compute time} taken strictly within our code logic (red line). The difference between the two is the platform's overhead for data transfers, coordination, scheduling tasks, etc.
	
	We observe that the \emph{weak-scaling} is inefficient. The graphs G20/P2, G30/P3 and G40/P4 have the same ratio of input load per unit compute resource ($\approx 10M$ vertices per VM). While the ideal weak-scaling times for these should be constant, instead they linearly increase. Doubling the resources between G40/4P to G40/8P does not halve the time taken, indicating that \emph{strong-scaling} is poor as well. As partition-count increases, the height of the merge-tree grows logarithmically. This increases the coordination costs for the partition-centric model, taking $2,3,3,4$ supersteps for $2,3,4,8$ partitions. So there is a trade-off between using less compute time against more cumulative data transfers and coordination costs.
	
	We do see that the user compute time (red line) is just half of the total time (blue line), and also grows at a slower rate. This indicates that Spark's distributed data transfer and I/O as part of its \emph{shuffle}, and the coordination costs for on-demand scheduling of tasks (e.g., 44 tasks are scheduled during the lifetime of G50/P8) and barrier synchronization at each application stage (e.g., 15 stages for G50/P8) dominate. They degenerate as the the graph size increases and the machines to transfer data across the commodity network increases.

	\begin{figure}[!t]
		\centering
		\includegraphics[width=0.85\columnwidth]{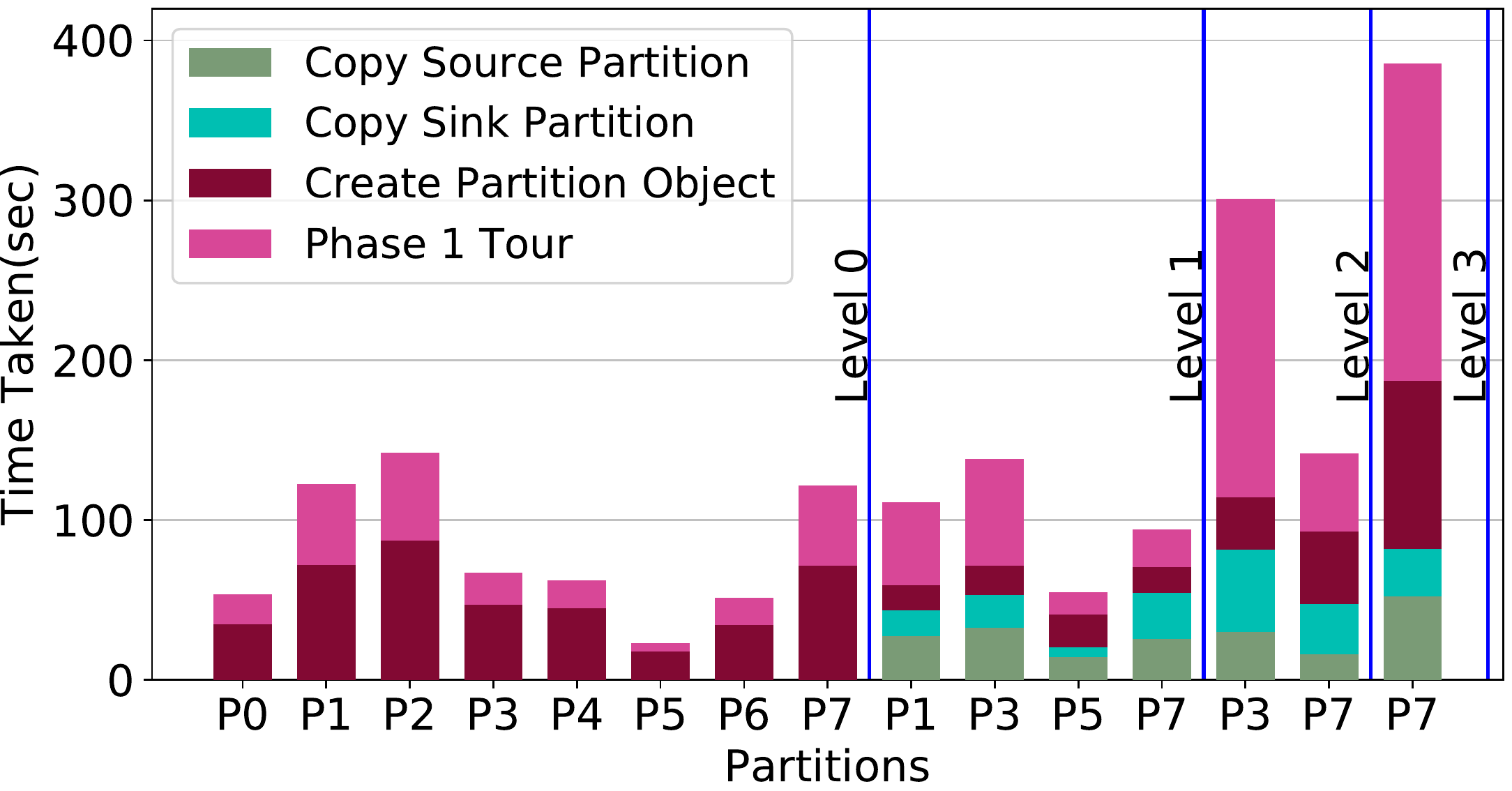}
		\caption{\emph{Split of user compute time} for each partition and level, for the G50/P8 graph}
		\label{plt:50m8pph1ph2time}
	\vspace{-0.1in}
	\end{figure}
	
	We further examine the \textbf{user compute time splits} for our application. Fig.~\ref{plt:50m8pph1ph2time} shows a stacked bar plot of various user times, spent in the Phase 1 computation (pink), and the Phase 2 partition merging (rest), for the \emph{G50/P8 graph}, across different partitions at different levels of the merge tree. In level 0 with 8 partitions, $P0-P7$, much of the time is spent in constructing the graph object from its storage format  
	(\emph{Create partition object}) with only $\approx 33\%$ of the time, on average, spent on the Phase 1 computation. Here, the graph sizes are large on disk ($\approx 605~MB$ per partition for G50/P8) and the I/O and Java object costs dominate. This continues in level 1 as well, with the additional cost of serializing, deserializing and merging the partition object pairs. In levels 2 and 3, each merged partition grows large enough that the compute time for Phase 1 starts to dominate with fewer partitions to move and merge, taking $\approx 48\%$ in level 2 and $\approx 51\%$ in level 3.
	We also observe that as the levels increase, the overall compute time increases as well. This shows that the time taken to process a merged partition is higher as more partitions are merged into one.
	
	These illustrate the challenges of scaling graph algorithms on distributed memory machines, even for algorithms like Euler circuits that have modest complexity. It also highlights the inefficiencies of Big Data platforms like Spark in handling graph data with irregular structure, with substantial time spent on data movement (albeit coarse-grained), and Java object management, compared to processing text or tuple data.
	
	\begin{figure}[t]
		\centering
		\subfloat[G40/P8 Graph\label{plt:40m8pComputeTime}]{\includegraphics[width=0.8\columnwidth]{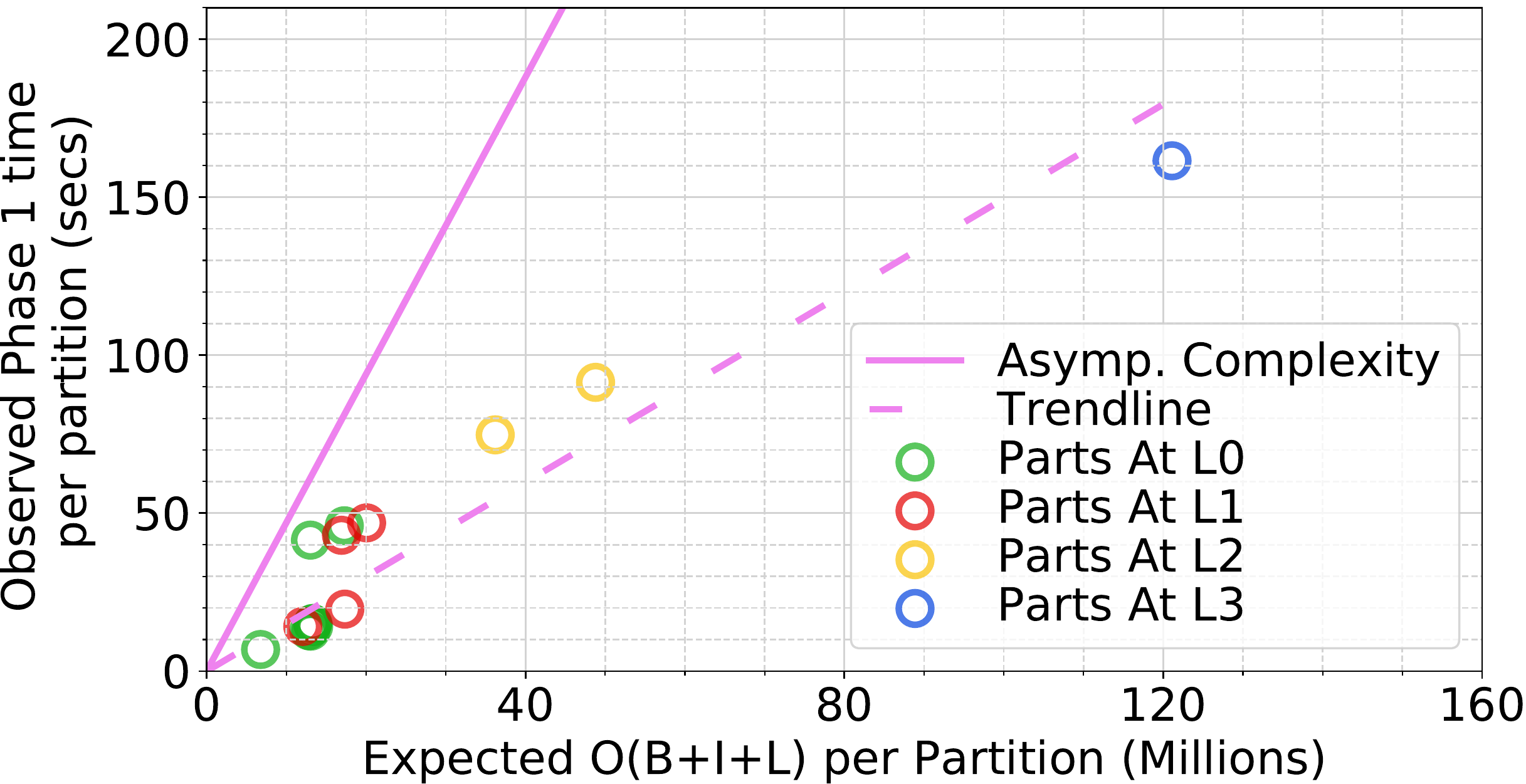}}\\
	\vspace{-0.1in}
		\subfloat[G50/P8 Graph\label{plt:50m8pComputeTime}] {\includegraphics[width=0.8\columnwidth]{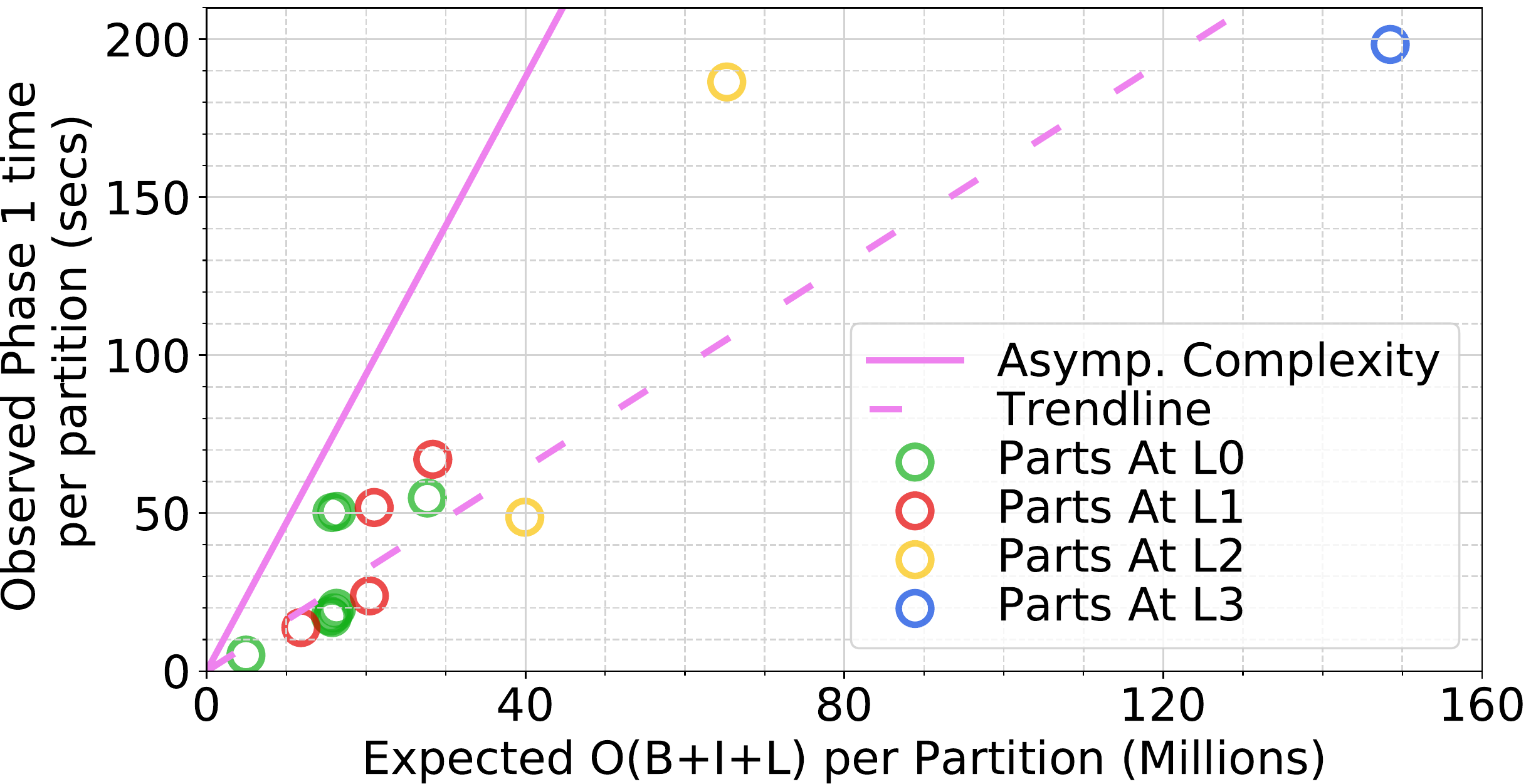}}
		\caption{\emph{Expected \& observed Phase 1 time}, per partition}
			\label{plt:ComputeTime}
	\vspace{-0.1in}
	\end{figure}

	Next, we specifically look at the \textbf{expected and observed times taken by Phase 1}, where the bulk of our actual graph traversal and circuit identification take place. Fig.~\ref{plt:ComputeTime} shows a scatter plot of the \emph{expected time complexity} of Phase 1 for each partition at each level, $\mathcal{O}(|B_i| + |I_i| + |L_i|)$, on the X axis, and the corresponding \emph{observed time in seconds}, on the Y axis. There are 8, 4, 2 and 1 partitions in levels 0, 1, 2 and 3, for a total of 15 data points. 
	We also show a dashed linear trend-line through these points, and a solid line with the asymptotic complexity, i.e., all observed values lie below this.
	
	We can clearly see that the expected time complexity closely matches the observed times. The slopes for both the G40/8P and G50/8P graphs are similar, and we can expect this to extrapolate to larger partitions and counts. So the computational cost for the critical Phase 1 algorithm is consistent with our design and analysis. We do see some outliers, e.g., in G50/8P's level 2. This is due to a skew in one of the two partitions at that level, as is seen in Fig.~\ref{plt:50m8pph1ph2time} for P3 at level 2. This is a consequence of the partition imbalances in level 0 caused by the external partitioner accumulating at higher levels.
	
	\begin{figure}[t]
		\centering
		\subfloat[G40/P8\label{plt:40m8pLoadObIdIm}]{\includegraphics[width=0.8\columnwidth]{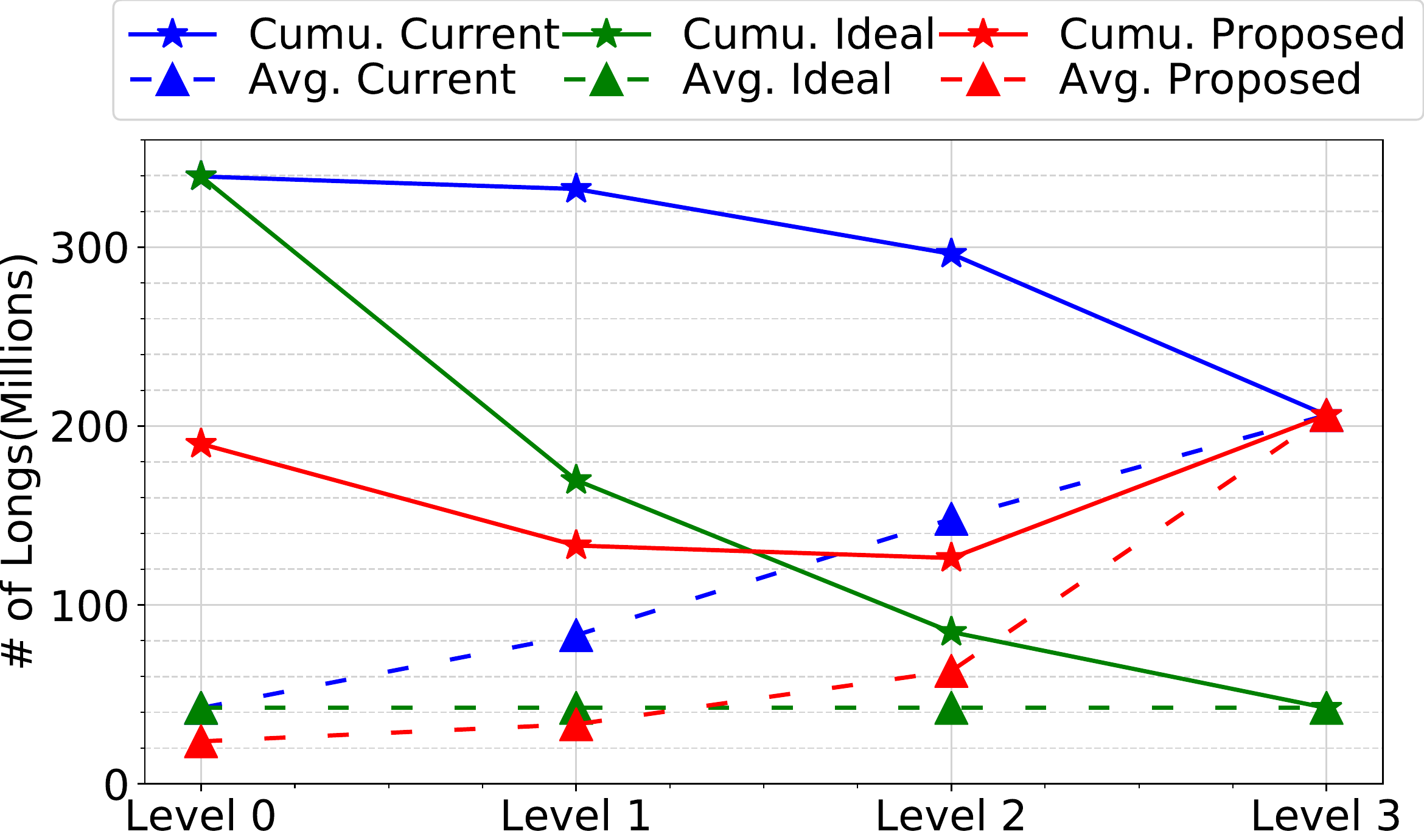}}\\
	\vspace{-0.1in}
		\subfloat[G50/P8\label{plt:50m8pLoadObIdIm}] {\includegraphics[width=0.8\columnwidth,trim={0 0 0 2cm},clip]{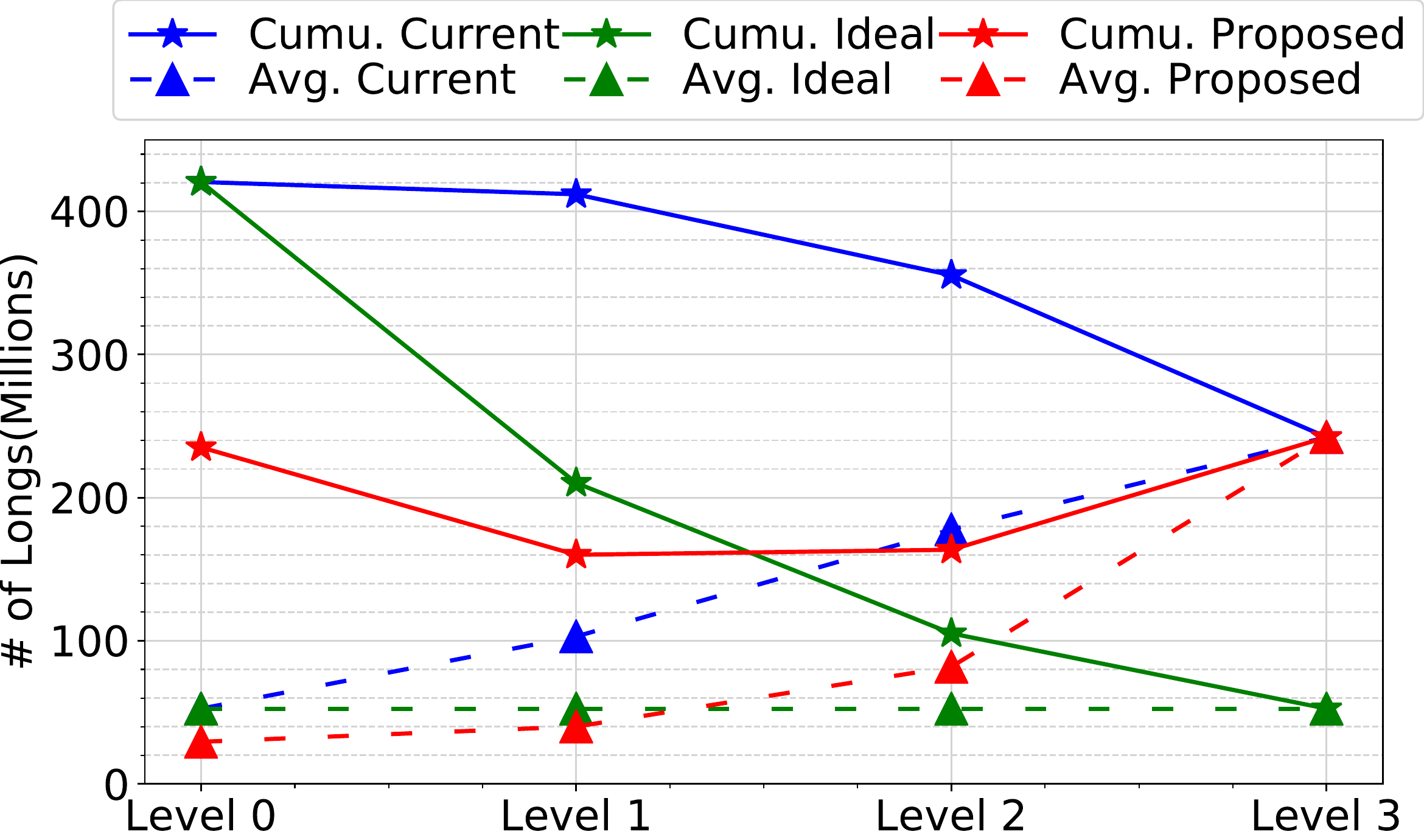}}
		\caption{\emph{Cumulative} and \emph{Average memory state size} for partitions at each level, for \emph{current, ideal} and \emph{proposed} approaches.} 
		  \label{plt:LoadObIdIm}
	\vspace{-0.1in}
	\end{figure}
	
	A key factor that limits our scaling is the \textbf{memory consumed} by the partitions at various levels. Each (merged) partition's state must fit in the memory of a single machine. Our algorithm design monotonically reduces the total in-memory state maintained across all partitions as we go up the level. This is done by replacing edge IDs of paths and cycles found in Phase 1 with a single path/cycle ID. But we also merge partitions up the level to allow remote edges to be included in the circuit, thus 
	causing the partition sizes to increase. In an \emph{ideal case}, the rate of drop in the state due to Phase 1 and the rate of increase in the state due to merging partitions should be identical, thus requiring a constant memory usage for partitions at any level. We study this behavior of the \emph{memory state} as the algorithm progresses. 
	
	Fig.~\ref{plt:LoadObIdIm} plots the \emph{cumulative (solid line) and average (dashed) number of \texttt{Int64}} (i.e., 8-byte \texttt{Long}) values that are maintained as part of the partitions' state at different levels. This is reported from within our user code, and is a platform-independent metric of the algorithm's memory use, compared to reporting the raw GB of RAM used as it is affected by the Java object structure, garbage collector, etc. We report this for our \emph{current algorithm} as blue lines. The cumulative is the distributed memory state at a level across all partitions, while the average is the per-machine (partition) size. We also report the \emph{ideal case} in green as a synthetic metric. Here, the state size for a merged partition matches the initial state size of its child partitions -- the average is constant across levels, while the cumulative is this times the number of partitions at a level.
	
	We see that for both the G40/P8 and G50/P8 graphs, there is a \emph{monotonic drop} in the cumulative state as the levels progress for our current algorithms, as expected. But the \emph{rate of drop is inadequate}. This is seen by the growth in the average partition size (blue dashed). While we expect the average size to be at $\approx 50M$ Longs, as seen at level 0, for G50/P8, this instead grows to $\approx 200M$ at level 3. Level 3 has just 1 merged partition rather than the 8 at level 0. But its size is $50\%$ of the cumulative size at level 0, rather than $12.5\%$ we expect.
	
	Also, the drop in cumulative size is more at higher levels rather than lower. In an ideal case, it should be exponential and drop more initially (concave green vs. convex blue solid lines). So the memory pressure is relieved only toward the last level. As a consequence, we are unable to run larger graphs with weak-scaling of memory as the required average memory increases with more levels.
	
	\begin{figure}[!t]
		\centering
		\includegraphics[width=0.8\columnwidth]{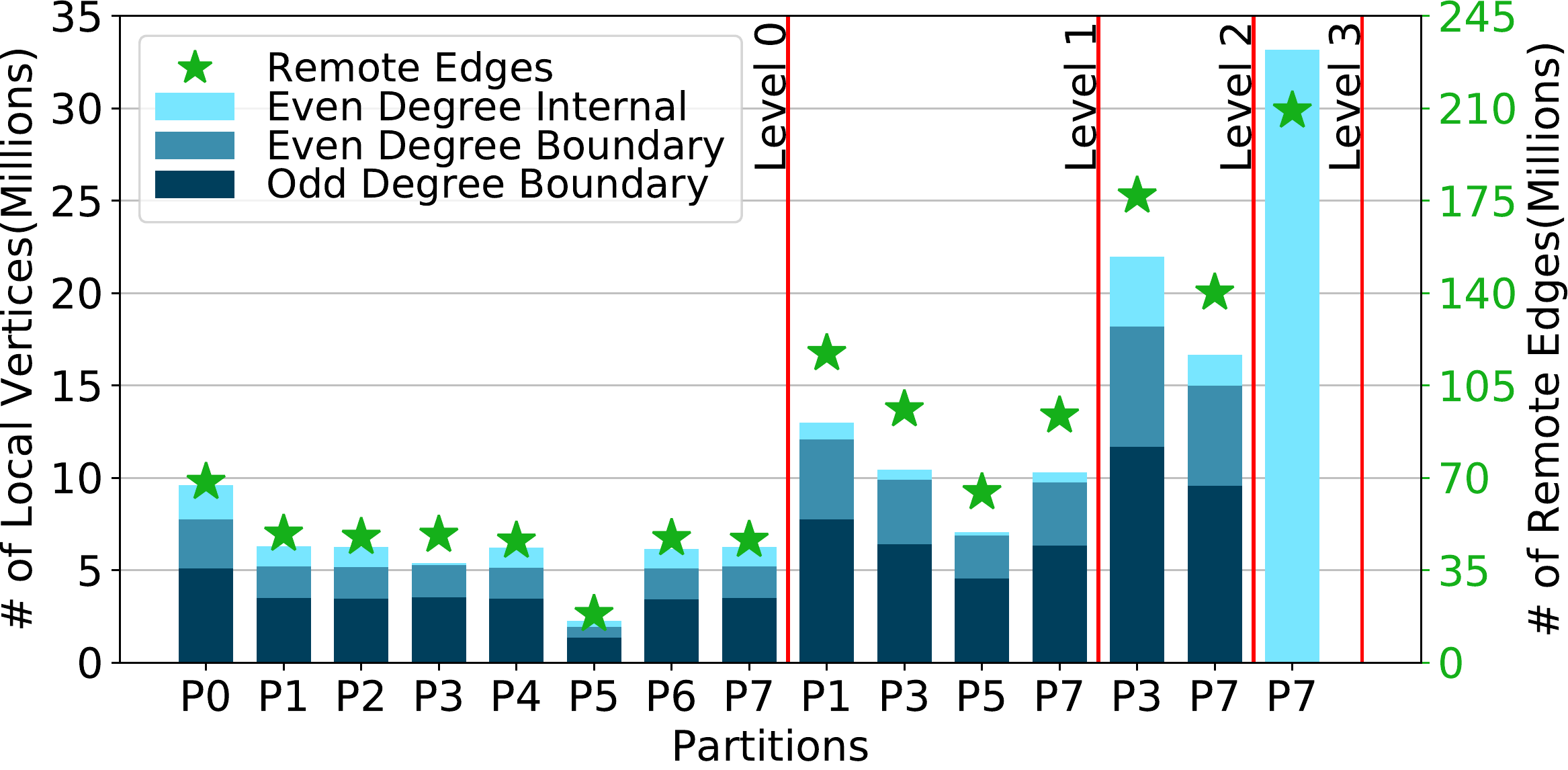}
		\caption{Vertices and edges per partition, for levels of G50/P8}
	\vspace{-0.15in}
		\label{plt:50m8pVertexDist}
	\end{figure}
	
	In Fig.~\ref{plt:50m8pVertexDist} (left Y axis), we examine the \textbf{number of vertices of different types} per-partition at \emph{the start of}
	Phase 1, for different levels of G50/P8, to analyze the drop rate. Boundary vertices have local and remote edges. Since only the local edges are consumed by Phase 1, the remote vertices are just carried over to the merged partition (right Y axis), with only some becoming local edges. The plots show that the number of boundary vertices and remote edges keep growing as we merge partitions. In particular, the remote edge count is $\approx 7\times$ the vertex count, dominates the memory usage.
	
	Lastly, we report that despite the use of efficient Java data structures, the overhead between the minimal raw bytes taken by the Longs and the actual memory used by the application is $\approx 10\times$. E.g.,  
	a partition that is  $950 MB$ on disk has a memory footprint of $12.8 GB$ within the Spark application.  
	At the root of the merge tree, the G50/8P partition consumes almost all
	of the JVM's allocated $45GB$ heap space. 
	
	\section{Proposed Improvements to Partition Merging}
	\label{sec:improved}

	This analysis shows that our proposed distributed algorithm for finding the Euler circuit is unable to weakly scale due to memory pressure. The drop rate of the partition sizes does not keep up with the growth rate due to the merging. This is primarily due to the remote edges in the merged partitions, as they accumulate up the levels of the merge tree. 
	
	In this section, we propose heuristics to reduce the number of remote edges that are maintained in memory to enable the algorithm to scale. We describe these techniques and analyze their potential benefits, but defer an implementation and empirical evaluation to future work.

	\para{Avoid Remote Edge Duplication} Our partitioned graph models undirected edges as a pair of directed edges, as is common to simplify traversals~\cite{giraph}. However, this doubles the memory usage for maintaining edges. As we see from Table~\ref{tab:dataset}, large graphs with many partitions have 70\% of their edges remote, and this duplication worsens the memory pressure.
	
	Intuitively, if two partitions are merging, it is sufficient for only one of them to hold a \emph{directed remote edge} between them, and convert it to a \emph{pair of directed internal edges} after they merge. This allows the other partition to drop its remote edges to this partition. E.g., in Fig.~\ref{fig:graph}, rather than P1 and P2 maintain the remote edges $e_{2,3}$ and $e_{3,2}$ to each other, respectively, only one needs to retain it, say $e_{2,3}$ in P1. When they merge into P2 in the next level, we convert the remote edge $e_{2,3}$ into a pair of internal edges $e_{2,3}$ and $e_{3,2}$ before Phase 1.
	
	The merge tree gives the partitions that will be merged at different levels, and this helps us decide at graph loading time on which partition should retain the remote edges among a pair that will (eventually) merge. Here, we select the partition that is \emph{heavier} among the pair, i.e., has more number of cumulative remote edges, as the one to drop its remote edges. This relieves its memory pressure, and also reduces the amount of state transfered from one level to another. Using this, the cumulative remote edge count kept in-memory will \emph{halve} at each level.
	
	\para{Defer Transfer of Remote Edges}
	In our current strategy, the entire $pathMap$ from a partition is transfered to a parent partition when they merge. This includes remote edges between the two partitions, and remote edges between the source partition to other partitions with which it will merge higher up the tree. E.g., in Fig.~\ref{fig:mergeTree}, when P1 and P2 merge into P2, P1 sends it its list of remote edges to P2, P3 and P4. While the edges to P2 become internal edges, the ones to P3 and P4 will only be used in level 2 but are in P2's memory in level 1. 
	
	In addition, we reduce the number of machines that are used to hold the merged partitions as we go up the merge-tree levels. So, while the total distributed memory on all machines is constant, the available distributed memory is halved at each level, even as the required memory does not drop as fast.
	
	Combining these two intuitions, we propose to delay copying of the remote edges from a child to an ancestor partition, until we are the level at which the ancestor is being merged. This reuses the machines on which the \emph{inactive leaf partitions} are present to hold the deferred remote edges, and sends the relevant remote edges to the \emph{active ancestor partition} at a higher level just before the latter's Phase 1 executes.

	In Fig.~\ref{fig:mergeTree}, the partitions P1--P4 retain their remote edges on their host machines' memory even after they complete level 0. Before level 1, P1 sends only its remote edges to P2 when it merges with it, but retains its remote edges to P3 and P4. Then, before level 2, P1 sends it remote edges to P3 and P4 to the merged ancestor P4, where they become internal edges are are used in P4's Phase 1. 
	As a result, we use memory available on machines hosting the leaf partitions, reduce memory pressure on the ancestors, and minimize the number of transfers of the the remote edges up the levels.
	
	\para{Analysis} We analytically model the impact of these two strategies on the memory usage of G40/8P and G50/8P in Fig.~\ref{plt:LoadObIdIm}, based on the previous experiments' traces. The dashed and solid red lines indicate the average and cumulative memory state expected. We see that the total memory state (solid red) in level 0 is $43\%$ smaller than the current approach (solid blue) due to avoiding holding the duplicate remote edges. As we saw in Fig.~\ref{plt:50m8pVertexDist}, the remote edge counts outstrip the vertex counts, and hence this steep drop.
	As the levels progress, we see the additional benefits from the deferred transfers. The total memory state drops more sharply from level 0 to levels 1 and 2, compared to the current approach. This is on top of the reduced base. The average state maintained in the active partitions (dashed red) grows slower in levels 1 and 2, and is $50$--$75\%$ smaller than the current approach (dashed blue). We are close to the ideal static value (dashed green); the reason the ``ideal'' is larger than the proposed for levels 0 and 1 is due to the avoiding remote edge duplication.
	Lastly, we note that our heuristics do not reduce the memory usage in the last level as there are no remote edges in this single merged partition. Mitigating this bottleneck is left to future work.
	
	\section{Conclusions}
	 \label{sec:conclusion}
	We have designed a partition-centric distributed algorithm for finding the Euler circuit over large graphs. Our algorithm finds local paths and cycles within each partition, and recursively merges them to form larger paths and cycles. These progressively reduce the memory used by replacing multiple edges with the path or cycle IDs they appear in. We analyze its complexity metrics, and evaluate its performance using Apache Spark. While the time complexity for the Phase 1 is as expected, the overheads data transfer in Spark's shuffle and object creation in Java reduce the benefits. Also, the memory pressure grows with the levels of merging due to remote edges that accumulate.
	We address this using heuristics to mitigate the growth in remote edge state and data transfers, and our analysis shows that these can reduce the memory usage by $50$--$75\%$. Its empirical validation is left to \emph{future work}, as is optimizing the last merged partition and evaluating the scalability for larger graphs. We will also consider generalizing this to non Eulerian graphs, by allowing edge revisits.
	
	\section{Acknowledgement}
	
	This work was supported by a Microsoft Data Science Fellowship provided to the first author, and a Microsoft Azure research grant for access to cloud resources.
	\bibliographystyle{IEEEtran}
	\bibliography{arxiv}
	
	\end{document}